\newif\ifconfver
\confvertrue        

\ifconfver
\documentclass[10pt,twocolumn,twoside]{IEEEtran}
\else
\documentclass[11pt,draftcls,onecolumn]{IEEEtran}
\fi

\usepackage{calc,amsfonts,amssymb,amsmath,bm,url,color,theorem,graphicx,cite}
\usepackage{psfrag,subfigure,float}
\usepackage{multirow,subfigure,amsmath,epsfig,amsfonts,amssymb,psfig,graphics,psfrag,theorem,calc,subfigure,url,bm,cite}
\usepackage{stfloats,hyperref}  
\usepackage{algorithm,algorithmic}
\usepackage{diagbox} 
\usepackage{hhline}
\usepackage{caption}
\usepackage{bbding}

\usepackage{mathtools}

\makeatletter
\def\multilimits@{\bgroup
	\Let@
	\restore@math@cr
	\default@tag
	\baselineskip\fontdimen10 \scriptfont\tw@
	\advance\baselineskip\fontdimen12 \scriptfont\tw@
	\lineskip\thr@@\fontdimen8 \scriptfont\thr@@
	\lineskiplimit\lineskip
	\vbox\bgroup\ialign\bgroup\hfil$\m@th\scriptstyle{##}$\hfil\crcr}
\def\Sb{_\multilimits@}
\def\endSb{\crcr\egroup\egroup\egroup}
\makeatother

{\begin{list}{}%
		{\setlength{\rightmargin}{0pc}%
			\setlength{\leftmargin}{1pc} }} %
	{\end{list}}

\newlength{\twidth}
\ifconfver
\else
\fi

\definecolor{orange}{RGB}{255,107,0}

\theorembodyfont{\rmfamily}

{\begin{list}{}{
			\settowidth{\labelwidth}{\mbox{\textnormal{#1}}}%
			\setlength{\leftmargin}{\labelwidth+\labelsep}}}%
	{\end{list}}

\definecolor{orange}{RGB}{255,107,0}

\ifconfver

\author{Yo-Yu Lai,~\IEEEmembership{Student Member,~IEEE}, Chia-Hsiang Lin,~\IEEEmembership{Member,~IEEE}, and \\ Zi-Chao Leng,~\IEEEmembership{Student Member,~IEEE}}

\title{
    Hyper-Restormer: A General Hyperspectral
	\\
	Image Restoration Transformer
    \\
    for Remote Sensing Imaging
 
    \thanks{This study was supported partly by the EINSTEIN Program of National Science and Technology Council (NSTC), Taiwan, under Grant MOST 111-2636-E-006-028; 
    and partly by the Emerging Young Scholar Program of NSTC, Taiwan, under Grant NSTC 112-2628-E-006-017.
    We thank the Center for Data Science at NCKU, and National Center for High-performance Computing (NCHC) for providing the computing resources.}

	\thanks{\textit{(Corresponding author: Chia-Hsiang Lin)}}
    \thanks{Y.-Y. Lai is with the Department of Electrical Engineering, National Cheng Kung University, Tainan, Taiwan (R.O.C.) 
    		(e-mail: q36104195@gs.ncku.edu.tw).}
	\thanks{C.-H. Lin is with the Department of Electrical Engineering, and with the Miin Wu School of Computing, National Cheng Kung University, Tainan, Taiwan (R.O.C.) 
		(e-mail: chiahsiang.steven.lin@gmail.com).}
	\thanks{Z.-C. Leng is with the Department of Electrical Engineering, National Cheng Kung University, Tainan, Taiwan (R.O.C.) 
		(e-mail: q38115558@gs.ncku.edu.tw).}
}

\else

\fi

\begin{document}

	\maketitle
	\ifconfver \else \vspace{-0.5cm}\fi

	\begin{abstract}
    The deep learning model Transformer has achieved remarkable success in the hyperspectral image (HSI) restoration tasks by leveraging Spectral and Spatial Self-Attention (SA) mechanisms. However, applying these designs to remote sensing (RS) HSI restoration tasks, which involve far more spectrums than typical HSI (e.g., ICVL dataset with 31 bands), presents challenges due to the enormous computational complexity of using Spectral and Spatial SA mechanisms. To address this problem, we proposed Hyper-Restormer, a lightweight and effective Transformer-based architecture for RS HSI restoration. First, we introduce a novel Lightweight Spectral-Spatial (LSS) Transformer Block that utilizes both Spectral and Spatial SA to capture long-range dependencies of input features map. Additionally, we employ a novel Lightweight Locally-enhanced Feed-Forward Network (LLFF) to further enhance local context information. Then, LSS Transformer Blocks construct a Single-stage Lightweight Spectral-Spatial Transformer (SLSST) that cleverly utilizes the low-rank property of RS HSI to decompose the feature maps into basis and abundance components, enabling Spectral and Spatial SA with low computational cost. Finally, the proposed Hyper-Restormer cascades several SLSSTs in a stepwise manner to progressively enhance the quality of RS HSI restoration from coarse to fine. Extensive experiments were conducted on various RS HSI restoration tasks, including denoising, inpainting, and super-resolution, demonstrating that the proposed Hyper-Restormer outperforms other state-of-the-art methods.

		\bfseries{\em Index Terms---}
		deep learning, 
        hyperspectral image,
        image restoration,
        remote sensing,
        Transformer.
        
	\end{abstract}

	\ifconfver \else \vspace{-0.0cm}\fi
	
	\ifconfver \else \vspace{-0.5cm}\fi
	
	\ifconfver \else  \fi

        \begin{figure}[h!]
		\centerline{\includegraphics[width=0.49\textwidth]{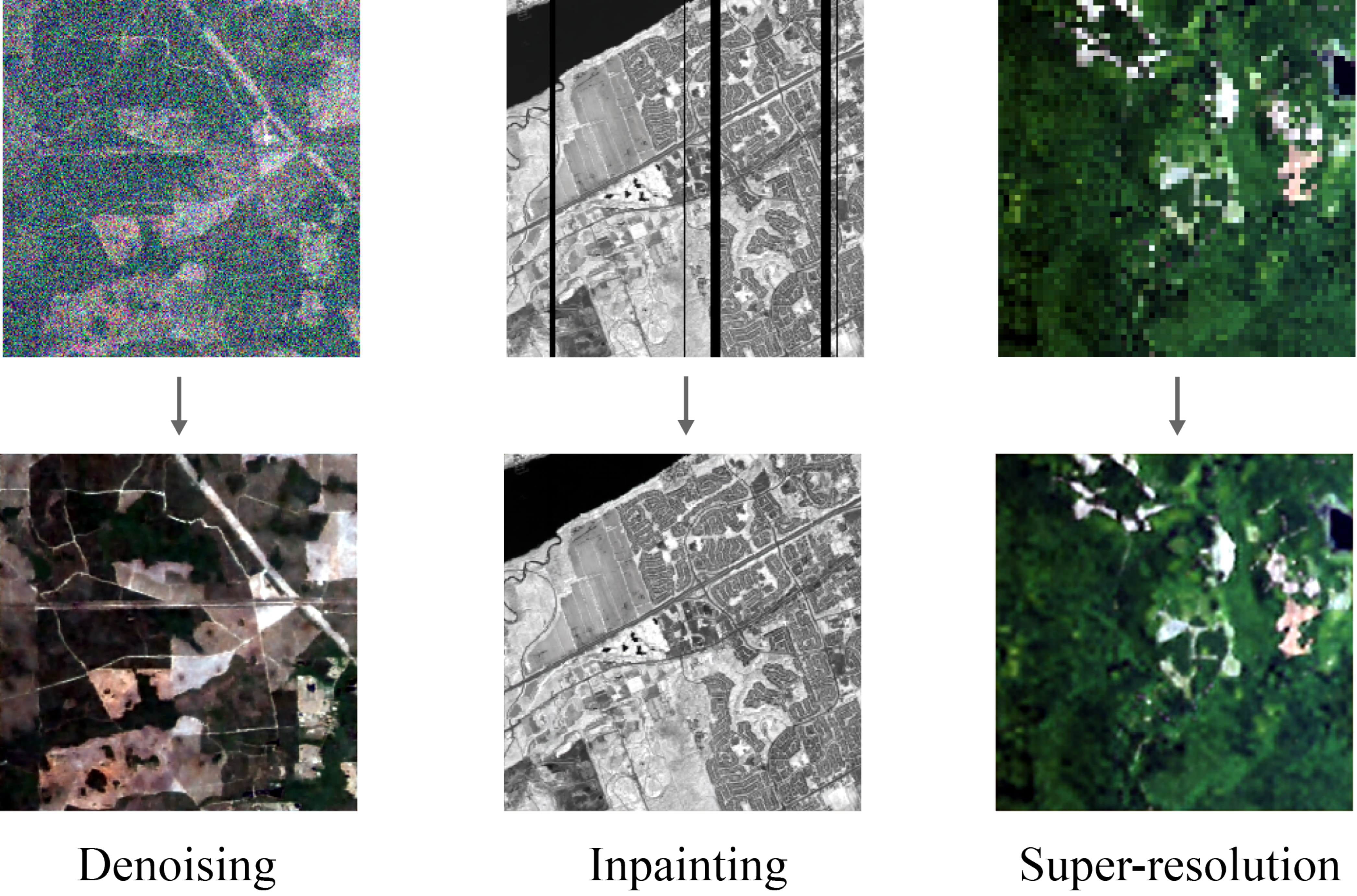}}
		\caption{
         Remote sensing HSIs with landscapes of farm, city, and vegetation respectively correspond to denoising, inpainting, and super-resolution HSI restoration tasks.
		}
		\label{fig:cover}
	\end{figure}
	
 
	\section{Introduction}\label{sec:introduction}
    Hyperspectral image (HSI)  provides a wealth of spectral information and are widely used for diverse applications, including earth observation, mineral exploration, environmental monitoring, and target detection. However, the quality of HSIs can be degraded by various factors, such as photon effects, atmospheric interference, and physical limitations of the sensors, resulting in issues such as random noise, stripe corruption, and low spatial resolution. These factors severely impact the usability of HSIs in those applications, making HSI restoration essential for enhancing image quality. Several HSI restoration tasks have been developed, including HSI denoising \cite{wang2017hyperspectral,zhuang2018fast, he2019non, sidorov2019deep, bodrito2021trainable, kan2021attention, peng2022fast, li2022spatial, yuan2018hyperspectral,wei20203, cao2021deep}, inpainting \cite{cerra2014unmixing, he2018hyperspectral, zhuang2018fast, sidorov2019deep, lin2021admm}, and super-resolution \cite{mei2017hyperspectral, li2018single, sidorov2019deep, li2020hyperspectral, jiang2020learning, lin2022single}. Most HSI restoration methods are developed for specific tasks, but a few can provide solutions for multiple HSI restoration problems, making them more widely applicable.

    Traditional HSI restoration methods typically rely on prior knowledge obtained from the images. For instance, the low-rank property \cite{wang2017hyperspectral, zhuang2018fast, he2019non, he2018hyperspectral} and sparse representation \cite{zhuang2018fast} are commonly used to reduce image noise and estimate missing information, respectively. The former represents the original image through a low-dimensional subspace, effectively removing noise, while the latter assumes that the image can be represented with fewer non-zero coefficients in a specific dictionary, allowing for the estimation of missing information. Additionally, total variation regularization \cite{wang2017hyperspectral, peng2022fast, he2018hyperspectral} is frequently employed to make the image sufficiently smooth and greatly reduce noise, while the non-local self-similarity method \cite{zhuang2018fast, he2019non} estimates each pixel's value by searching for similar regions in the HSI, which are usually located at other positions. Despite their effectiveness in HSI restoration, these methods often require manual parameter tuning and may not perform well in complex real-world scenarios.

    With the development of deep learning, Convolutional Neural Networks (CNNs) have become a solution for various tasks in HSI restoration, such as denoising \cite{yuan2018hyperspectral, sidorov2019deep, kan2021attention}, fusion \cite{han2018ssf}, and super-resolution \cite{li2017hyperspectral, sidorov2019deep}. Unlike traditional methods, CNN-based approaches do not rely on manually designed priors but require the appropriate model structure, enabling CNN to learn suitable and effective feature representations from data. Moreover, with the growth of big data and the improvement of hardware techniques, deep learning-based methods have outperformed most traditional image restoration methods. As a result, CNN-based methods have been widely employed in the field of hyperspectral imaging and have exhibited remarkable results. However, the typical 2-D CNN performs convolution only in the spatial domain, which cannot effectively capture the correlation between spectral bands. For HSIs with high spectral similarity between bands, leveraging spectral information can substantially enhance the restoration performance. Therefore, 3-D CNN-based methods \cite{mei2017hyperspectral, li2020hyperspectral, wei20203} have emerged for HSI restoration to overcome the limitation of the typical 2-D CNN. 3-D CNN can simultaneously capture both spatial and spectral information, making it possible to effectively restore HSIs with high spectral similarity between bands.
    
    Recently, the Transformer architecture \cite{vaswani2017attention}, originally developed for natural language processing, has been adapted to computer vision tasks. Transformer-based structures use global self-attention mechanisms to capture long-range dependencies in the feature maps, overcoming the limitations of CNNs in obtaining non-local information. It has led to significant achievements in computer vision, including various HSI-related tasks.
    Nevertheless, the utilization of global self-attention computation has led to quadratic computational cost, resulting in significant computational complexity for vision applications. To tackle this issue, the Swin Transformer \cite{liu2021swin} was introduced, which utilizes Window-based Self-Attention and Shift-windows mechanisms to significantly reduce computational complexity, achieving remarkable performance in image classification. 
    Moreover, Uformer \cite{wang2022uformer} and Restormer \cite{zamir2022restormer}, employ Window-based self-attention combined with U-shaped hierarchical model structure design to reduce computational complexity while achieving outstanding performance in image restoration. However, these self-attention mechanisms are designed for spatial correlation and cannot effectively utilize the spectral correlation of HSIs. To address this issue,  MST \cite{cai2022mask} and MST++ \cite{cai2022mst++} were later developed, which employ a Spectral-wise Self-Attention mechanism to effectively obtain global information among spectral bands for HSI spectral reconstruction. To better address HSI restoration, some \cite{li2022spatial, yu2023dstrans, lai2023mixed} have started to utilize both Spectral Self-Attention and Spatial Self-Attention mechanisms. These models can better capture the long-range dependencies between spectral and spatial dimensions, leading to improved restoration performance.

    However, the recently proposed state-of-the-art deep learning-based HSI restoration methods that use novel mechanisms such as 3-D CNNs or self-attention, most of them are tailored for 31-band HSI datasets (e.g., ICVL \cite{arad2016sparse}, CAVE \cite{yasuma2010generalized}, and  Harvard \cite{chakrabarti2011statistics}). Applying these methods to remote sensing HSIs with much more spectral bands, they often encounter GPU out-of-memory issues during the training stage due to the massive parameters and computational requirements. As a result, some methods can only use pre-trained weights from other datasets or cut the data into smaller pieces during the training of remote sensing HSI restoration, which prevents them from adequately learning the unique characteristics of remote sensing HSIs.

    In this paper, we propose Hyper-Restormer, a lightweight and effective Transformer-based architecture for remote sensing HSI restoration, which can be used for denoising, inpainting, and super-resolution tasks. First, we propose a novel Lightweight Spectral-Spatial (LSS) Transformer Block that utilizes both Spectral Self-Attention and Spatial Self-Attention mechanisms to capture long-range dependencies in spectral and spatial domains. We also propose a novel Lightweight Locally-enhanced Feed-Forward Network (LLFF) to enhance local content information without requiring an excessive computational cost. LSS Transformer Blocks are combined to form a Single-stage Lightweight Spectral-Spatial Transformer (SLSST). The SLSST's novel model structure is to efficiently exploit the low-rank property of HSIs by decomposing the input feature maps into basis and abundance components. After the decomposition, the number of parameters and the size of the feature maps can be greatly reduced, thereby significantly reducing the computational complexity associated with using Spectral Self-Attention and Spatial Self-Attention mechanisms. Finally, multiple SLSSTs are cascaded to form Hyper-Restormer, which utilizes a multi-stage restoration strategy to restore HSIs from coarse to fine levels. 

    Overall, we summarize the contributions of this paper as follows:
    \begin{itemize}
			\item
            We propose a novel framework, Hyper-Restormer, for various remote sensing HSI restoration tasks.
   
			\item
            We propose a novel model structure that conforms to the low-rank property of HSIs, designed to significantly reduce the computational complexity of using self-attention mechanisms. 

            \item
            We propose a novel Lightweight Locally-enhanced Feed-Forward Network, which enhances local context information with lightweight computational cost.

            \item
            Extensive experiments were conducted on both simulated and real remote sensing HSI data, demonstrating that the proposed Hyper-Restormer framework outperforms other state-of-the-art methods in various HSI restoration tasks.
   
	\end{itemize}

    In the remaining sections of this article, we organize the content as follows. In Section \ref{sec:related_work}, we briefly review various methods for remote sensing HSI restoration tasks. In Section \ref{sec:proposed_method}, we present our proposed remote sensing HSI restoration method, Hyper-Restormer, by introducing the model architecture in a top-down manner, gradually delving into the details of each module. In Section \ref{sec:experiment}, we conduct extensive experiments on simulation and real HSIs to demonstrate the superiority of Hyper-Restormer. Moreover, we perform ablation studies to validate the effectiveness of the proposed modules. Finally, we summarize the conclusions in Section \ref{sec:conclusion}.
    
    \section{Related Works}\label{sec:related_work}
    In this section, we briefly review several recent methods for remote sensing HSI restoration compared in the paper. Specifically, we will cover the research related to HSI denoising in Section \ref{sec:denoise_peer}, HSI inpainting in Section \ref{sec:inpaint_peer}, and HSI super-resolution in Section \ref{sec:sr_peer}, respectively.
    
    \subsection{Remote Sensing HSI Denoising Methods}\label{sec:denoise_peer}
    For the denoising task, the main goal is to remove noise from the image. 
    The low-rank property and total variation regularizers are frequently used in optimized-based HSI denoising methods.
    LRTDTV \cite{wang2017hyperspectral} leverages spatial-spectral total variation to ensure that the restored image is sufficiently smooth in both spatial and spectral domains, coupled with the low-rank property to significantly reduce noise in HSIs. Another commonly used property is sparse representations. 
    FastHyDe \cite{zhuang2018fast} decomposes the HSI by low-rank property and then denoises through the sparse representations and non-local similarity method BM3D \cite{bm3d}. 
    Similarly, NGmeet \cite{he2019non} employs low-rank property and non-local similarity to jointly learn and update the orthogonal basis and reduced image for HSI denoising. 
    In addition to traditional optimization-based methods, deep learning-based methods have rapidly developed for hyperspectral denoising tasks.
    Unsupervised deep learning-based method DHP \cite{sidorov2019deep} applies the concept of the deep image prior \cite{ulyanov2018deep} to HSIs, using the network decoder structure as intrinsic image priors for HSI denoising.
    T3SC \cite{bodrito2021trainable} proposes a hybrid method based on sparse coding principles but parameterizes the entire optimization process by end-to-end model training. Hence, the method retains the interpretability of the deep learning model.
    AODN \cite{kan2021attention} uses multiscale separable convolution to explore adjacent spatial-spectral information and reduce model complexity. Furthermore, it suppresses noise through an Octave kernel and attention mechanism.
    Fast-optimized-based algorithms have become increasingly popular in optimized-based methods. 
    RCTV \cite{peng2022fast} proposes a representative coefficient total variation regularizer, which can simultaneously capture the low-rank and local smooth properties. With this low-computational-complexity regularizer, it achieves comparable speeds to deep learning-based methods. 
    The Transformer model, which utilizes self-attention mechanisms, has become the most popular deep learning-based method in recent years. 
    SST \cite{li2022spatial} utilizes non-local spatial self-attention and global spectral self-attention to capture similarity characteristics in both the spatial and spectral dimensions, achieving excellent HSI denoising performance.

    \subsection{Remote Sensing HSI Inpainting Methods}\label{sec:inpaint_peer}
    The objective of the inpainting task is to restore missing stripes caused by a damaged or aging sensor array. Interpolation is a simple and fast method for filling in missing values. 3D-PDE \cite{d2008inpainting} utilizes the surrounding known pixel area to restore the missing pixels. There are currently many inpainting methods that are optimized-based. These methods usually transform the inpainting problem into an optimization problem and recover the missing values by designing appropriate objective functions and constraints. 
    UBD \cite{cerra2014unmixing} transforms the HSI inpainting problem into a HSI unmixing problem and assumes that pure pixels exist in the HSI.
    The low-rank property of HSIs is also frequently utilized.
    LLRSSTV \cite{he2018hyperspectral} uses a spatial-spectral total variation regularization to ensure sufficient smoothness between spatial and spectral dimensions, coupled with the low-rank property for HSI inpainting.
    Similar to FastHyDe, FastHyIn utilizes HSI self-similarity and low-rank property to recover the missing pixels.
    Recently, deep learning methods have been employed in inpainting tasks.
    DHP \cite{sidorov2019deep} incorporates a masking mechanism in the model learning criterion to restore the missing pixels in the HSI.
    ADMM-ADAM \cite{lin2021admm} combines the advantages of convex optimization and deep learning by introducing a simple Q-norm regularizer. Believe that the preliminary inpainting result obtained from the deep learning model contains crucial information. Hence, the regularizer fused the information into the final result to improve the restoration quality.

    \subsection{Remote Sensing HSI Super-resolution Methods}\label{sec:sr_peer}
    The super-resolution task aims to recover high spatial resolution images from low spatial resolution ones. Recently, deep learning methods have achieved tremendous success in HSI super-resolution.
    3D-FCNN \cite{mei2017hyperspectral} utilizes 3-D convolution to extract information from both spatial and spectral dimensions, addressing the issue of typical 2-D convolution having a poorer ability to capture inter-spectral correlations.
    GDRRN \cite{li2018single} employs a grouped recursive module to transform the input HSI. Additionally, it combines the mean squared error loss and spectral angle mapper loss in training to improve the quality of the results and prevent spectral distortion.
    DHP \cite{sidorov2019deep} can also be used for super-resolution by modifying the learning criterion with an additional downsampling operation, demonstrating the versatility of the deep image prior apply in DHP.
    3D-GAN \cite{li2020hyperspectral} utilizes a 3-D convolutional generative adversarial network framework to generate high spatial resolution HSIs while incorporating spatial-spectral constraints in loss function to mitigate spectral distortion and texture blur.
    SSPSR \cite{jiang2020learning} utilizes spatial-spectral blocks to capture both spatial and spectral information in HSIs. The network also utilizes group convolution with shared weights to stabilize the training process.
    ADMM-Adam SR \cite{lin2022single} is designed based on the ADMM-Adam theory \cite{lin2021admm}. It utilizes a pre-trained neural network to obtain upsampled hyperspectral eigenimage. The upsampled eigenimage contains information that benefits the super-resolution task. The information is fused into the final result by a simple regularizer.

    \begin{figure*}[t]
		\centerline{\includegraphics[width=1.03\textwidth]{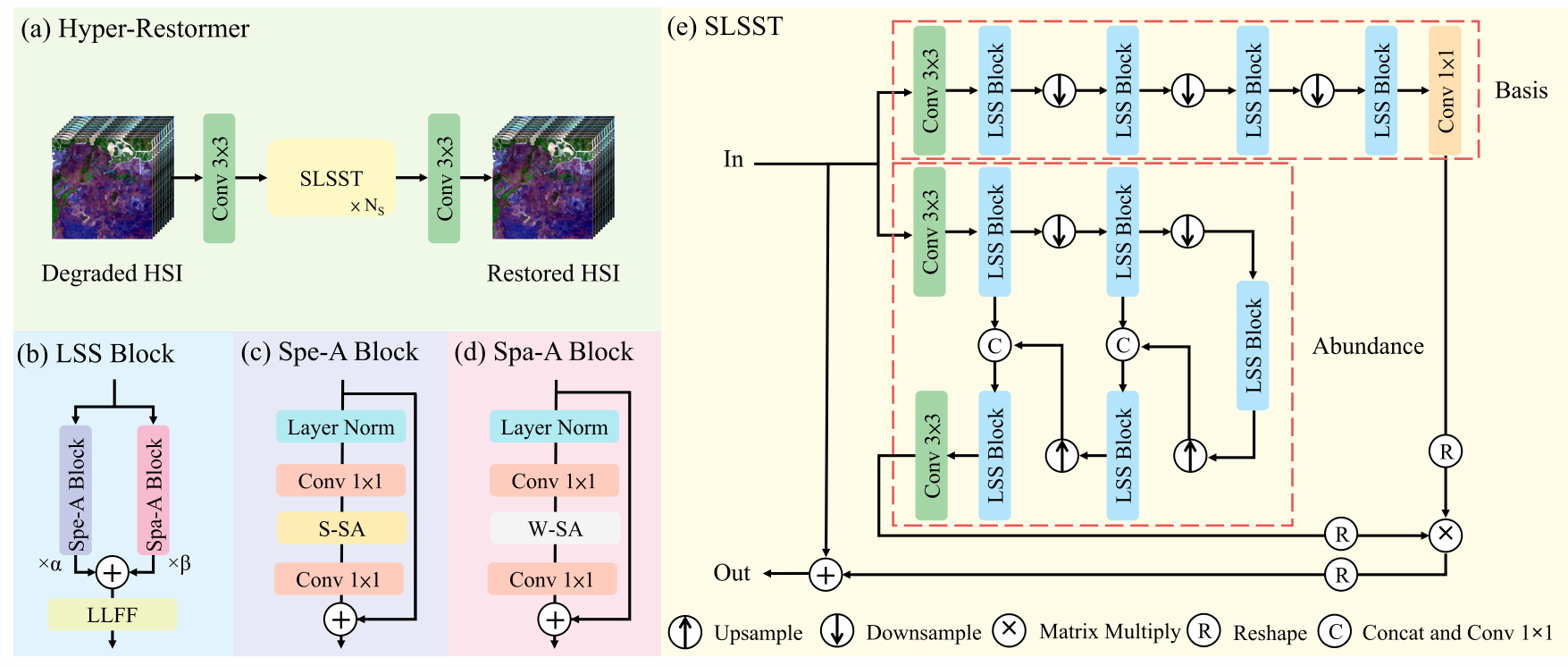}}
		\caption{{The overall pipeline of Hyper-Restormer. (a) Hyper-Restormer. (b) Lightweight Spectral-Spatial Transformer Block. (c) Spectral Attention Block. (d) Spatial Attention Block. (e) Single-stage lightweight Spectral-Spatial Transformer.}}
		\label{fig:framework}
    \end{figure*}

    \section{Proposed Method}\label{sec:proposed_method}
    In this section, we introduce the proposed HSI restoration method Hyper-Restormer. First, we describe the complete process and model structure of Hyper-Restormer (cf. Section \ref{sec:pipeline}). Then, we introduce Single-stage Lightweight Spectral-Spatial Transformer (SLSST), which builds up Hyper-Restormer, and its novel low-rank structural design (cf. Section \ref{sec:slsst}).  Finally, we present Lightweight Spectral-Spatial (LSS) Transformer Block, the fundamental component of the SLSST, including the self-attention mechanism and the proposed Lightweight Locally-enhanced Feed-Forward Network (LLFF) used within the LSS Transformer Block (cf. Section \ref{sec:lss}).


    \subsection{Overall Pipeline} \label{sec:pipeline}
    The proposed Hyper-Restormer consists of $N_S$ cascaded SLSSTs, as shown in Figure \ref{fig:framework}(a). Given a degraded HSI $\mathbf{D} \in \mathbb{R}^{C \times H \times W}$, with channels $C$, height $H$, and width $W$, respectively. Initially, Hyper-Restormer applies a 3 × 3 convolutional layer to extract low-level features $\mathbf{F}_0 \in \mathbb{R}^{E \times H \times W}$, where $E$ is the embedding dimension. Then, $N_S$ SLSSTs are sequentially applied to restore the HSI from coarse to fine. Finally, another 3 × 3 convolutional layer projects the final output back to the original dimension $C$, obtaining the restoration result $\mathbf{R} \in \mathbb{R}^{C \times H \times W}$. The overall HSI restoration process is represented as:
    \begin{equation}
    \begin{aligned}
        \mathbf{F}_0 = \mathbf{Conv}(\mathbf{D}), \\
        \mathbf{F}_{s} = \mathbf{SLSST}(\mathbf{F}_{s-1}), s=1,2,......N_{S},\\
        \mathbf{R} = \mathbf{Conv}(\mathbf{F}_{N_S}), \\
    \end{aligned}
    \end{equation}
    where $s$ denotes the stage of the SLSST block.
    
    Figure \ref{fig:framework}(e) depicts the SLSST, composed of LSS Transformer Blocks that leverage the Spectral and Spatial Self-Attention mechanism to capture long-range dependencies while reducing computational cost through specially designed low-rank model architecture. The SLSST comprises a sequential basis module and a U-shaped abundance module. They are used to generate basis component $\mathbf{B}_{s} \in \mathbb{R}^{E \times \sqrt{N_{B}} \times \sqrt{N_{B}}}$, and abundance component $\mathbf{A}_{s} \in \mathbb{R}^{N_B \times H \times W}$, respectively, where $N_B$ represents the number of basis chosen. The outputs are reshaped and then multiplied together before being added with the residual connection, obtaining the final output of the block $\mathbf{F_s} \in \mathbb{R}^{E \times H \times W}$. The computation process in SLSST could be denoted as follows:
    \begin{equation}
    \begin{aligned}
        \mathbf{B}_{s} = \mathbf{Basis \, Module}(\mathbf{F}_{s-1}) \\
        \mathbf{A}_{s} = \mathbf{Abundance \, Module}(\mathbf{F}_{s-1}), \\
        \mathbf{B'}_{s},\mathbf{A'}_{s} = \mathbf{Reshape}(\mathbf{B}_{s}), \mathbf{Reshape}(\mathbf{A}_{s})\\
        \mathbf{F}_{s} = \mathbf{F}_{s-1} + \mathbf{Reshape}(\mathbf{B'}_{s}\mathbf{A'}_{s}).
    \end{aligned}
    \end{equation}

    To avoid excessive model parameters that can result from using traditional convolution on a large number of channels, Hyper-Restormer utilizes a 4 × 4 Depthwise-Separable Convolution \cite{howard2017mobilenets} with stride 4 for the downsampling operation by a factor of 4. The upsample operation, on the other hand, is achieved through the use of pixel shuffle \cite{shi2016real} with a 3 × 3 convolutional kernel.

    \subsection{Single-stage Lightweight Spectral-Spatial Transformer (SLSST)} \label{sec:slsst}
    The SLSST can be decomposed into two parts: a sequential basis module and a U-shaped abundance module. In each module, input feature maps $\mathbf{F}_{s-1} \in \mathbb{R}^{E \times H \times W}$ are first projected into the designated channel dimensions $\mathbf{F}_{B, \, s} \in \mathbb{R}^{E \times H \times W}$ and $\mathbf{F}_{A, \, s} \in \mathbb{R}^{N_B \times H \times W}$, respectively. Subsequently, both $\mathbf{F}_{B, \, s}$ and $\mathbf{F}_{A, \, s}$ undergo repeated processing through LSS Transformer Blocks to capture long-range dependencies and downsampling layers to reduce its spatial dimension. 

    The sequential basis module repeats the process until the spatial dimensions reach the desired values, resulting in the product of length and width equal to $N_B$. The U-shaped abundance module, on the other hand, follows a standard U-shaped model architecture, where the channel dimension is doubled after downsampling until it reaches the bottleneck. Then channel dimension is halved after upsampling until it returns to the original dimension. This part utilizes a U-shaped structure with skip connections to capture multi-resolution contextual information.

    Before producing the final output, the feature maps from the sequential basis module go through a 1 × 1 convolutional layer to enhance the correlation between channels, while the U-shaped abundance module is processed through a 3 × 3 convolutional layer to enhance local information. Finally, the output feature maps from both parts $\mathbf{B}_{s} \in \mathbb{R}^{E \times \sqrt{N_B} \times \sqrt{N_B}}, \mathbf{A}_{s} \in \mathbb{R}^{N_B \times H \times W}$ are reshaped into specific dimensions $\mathbf{B'}_{s} \in \mathbb{R}^{E \times N_B}, \mathbf{A'}_{s} \in \mathbb{R}^{N_B \times HW}$ and multiplied them to obtain result $\mathbf{B'}_{s}\mathbf{A'}_{s} \in \mathbb{R}^{E \times HW}$. The result is then reshaped, added with a residual connection, and yields the output of the SLSST block $\mathbf{F}_s \in \mathbb{R}^{E \times H \times W}$. 
    
    The following is a more detailed explanation of the design of the low-rank model architecture.

    \begin{figure*}[t]
		\centerline{\includegraphics[width=1.03\textwidth]{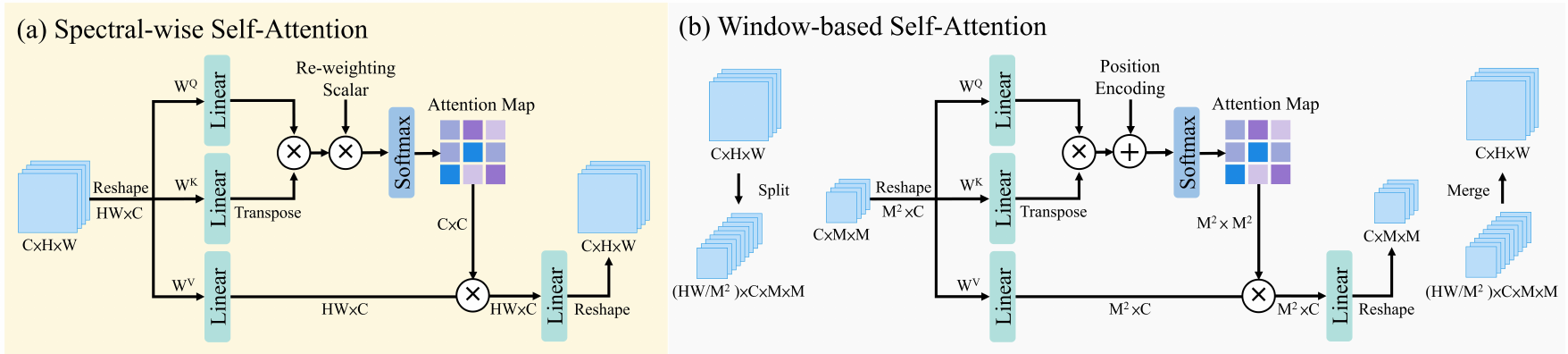}}
		\caption{{Illustration of (a) Spectral-wise Self-Attention and (b) Window-based Self-Attention.}}
		\label{fig:self-attention}
    \end{figure*}

    \subsubsection*{Low-rank Model Architecture}

   Low-rank is a fundamental property of the HSI, which means that the spectral vectors of the HSI data exist in low-dimensional subspaces \cite{bioucas2012hyperspectral}. By leveraging this property, the HSI can be decomposed into an orthogonal basis multiplied by non-negative abundance coefficients that sum to one \cite{lin2015fast}, often used in optimization problems to reduce a significant amount of computation. Based on this concept, we aim to split the design of the SLSST into two parts, one for generating the basis component and the other for generating the abundance component. Finally, we multiply the results of both parts to obtain the output.


    The U-shaped architecture is a commonly adopted approach for capturing multi-resolution contextual information. However, this approach doubles the channels after each downsampling operation, which can result in a significant computational burden when performing self-attention or convolution computation on hyperspectral images with a high number of channels. As a result, it is challenging to apply the U-shaped architecture to hyperspectral images.


    Optimization-based methods often simplify the original learning criterion of HSI into the abundance component and perform computations directly on the abundance component to reduce computational cost and stabilize the optimization process. They then multiply the result by the basis component to return to the original dimension. Inspired by this approach, we apply the computationally expensive U-shaped architecture to the abundance component, where the number of abundance channels $N_B$ is much less than the input feature maps embedding dimension $E$. We combine the U-shaped architecture abundance component with a sequential architecture basis component to generate the final output by multiplying the two components. By adopting this design, SLSST can significantly reduce the computational burden while capturing multi-resolution contextual information.
    

    \subsection{Lightweight Spectral-Spatial (LSS) Transformer Block} \label{sec:lss} 
     Figure \ref{fig:framework}(b) illustrates the components of the LSS Transformer Block, including the Spectral Attention Block, Spatial Attention Block, and LLFF.
    In the LSS Transformer Block, the input feature maps are passed through parallel arranged Spatial Attention Block and Spectral Attention Block to capture long-range dependencies along spectral and spatial dimensions, respectively. The outcome of each block is multiplied by learnable reweighting scalar before passing through the LLFF to enhance local information, resulting in the final output of the LSS Transformer Block. The overall process of the LSS Transformer Block could be denoted: 
    \begin{equation}
    \begin{aligned}
        \mathbf{F}_{Spe} = \mathbf{Spectral \, Attention}(\mathbf{F}_{LSS_0}), \\
        \mathbf{F}_{Spa} = \mathbf{Spatial \, Attention}(\mathbf{F}_{LSS_0}), \\
        \mathbf{F}_{LSS} = \mathbf{LLFF}(\alpha \mathbf{F}_{Spe} + \beta \mathbf{F}_{Spa} ),
    \end{aligned}
    \end{equation}
     where $\mathbf{F}_{LSS_0}$ denotes the input of LSS Block and $\mathbf{F}_{LSS}$ denotes the output of LSS Transformer Block. $\alpha, \beta$ are learnable reweighting scalars.
    
    The self-attention mechanisms used within the blocks and the proposed LLFF are described in detail as follows.
    
    \subsubsection{Spectral Attention (Spe-A) Block and Spatial Attention (Spa-A) Block} 
    HSI exhibits a high degree of similarity between spectral bands. Effectively utilizing this property can help with HSI restoration tasks.
    In addition, non-local self-similarity in the spatial domain has been extensively used in image restoration tasks. 
    Therefore, in HSI restoration, it is necessary to effectively utilize spectral and spatial information to improve the restoration performance.

    To capture both spectral and spatial long-range dependence in the HSI, Spectral-wise Self-Attention (S-SA) and Window-based Self-Attention (W-SA) mechanisms are employed in Spe-A Block and Spa-A Block, respectively. S-SA captures global information across the spectral bands, while W-SA captures global information across the spatial dimension. 
    
    In these blocks, the input feature maps are first normalized using layer normalization and then projected to low-dimensional subspace through a 1 × 1 convolution to reduce computational complexity. S-SA and W-SA are then applied to capture global information across spectral and spatial dimensions, respectively. Finally, the output is obtained by projecting back to the original dimension using another 1 × 1 convolution and adding with a residual connection.

    The following are computational equations regarding the S-SA and W-SA mechanisms. Details of S-SA and W-SA are also given in Fig. \ref{fig:self-attention}(a) and (b).

    \paragraph{Spectral-wise Self-Attention (S-SA)}
    The S-SA is designed to capture spectral long-range dependencies in the input tensor. To apply the S-SA, we first reshape and transpose the input tensor $\mathbf{X}_{in} \in \mathbb{R}^{C \times H \times W}$ to obtain $\mathbf{X} \in \mathbb{R}^{HW \times C}$. Then, we define projection matrices $\mathbf{W^Q}, \mathbf{W^K}, \mathbf{W^V}$ with size $\mathbb{R}^{C \times C}$ for queries, keys, and values, respectively. These projection matrices are used to obtain the query $\mathbf{Q}$, the key $\mathbf{K}$, and the value $\mathbf{V}$ as $\mathbf{Q}=\mathbf{X}\mathbf{W^Q}$, $\mathbf{K}=\mathbf{X}\mathbf{W^K}$, and $\mathbf{V}=\mathbf{X}\mathbf{W^V}$. The S-SA mechanism can be defined as follows:

    \begin{equation}
    \begin{aligned}
    \mathbf{S \textendash SA}(\mathbf{Q}, \mathbf{K}, \mathbf{V}) = \mathbf{V} \times \mathbf{Softmax}(\sigma \mathbf{K^T}\mathbf{Q}),
    \end{aligned}
    \end{equation}
    
    where the learnable parameter $\sigma$ is a re-weighting scalar, and $\times$ represents matrix multiplication.

    \paragraph{Window-based Self-Attention (W-SA)}
    On the other hand, the W-SA is designed to capture spatial long-range dependencies in the input tensor. To apply the W-SA, we first split the input tensor $\mathbf{X}_{in} \in \mathbb{R}^{C \times H \times W}$ into non-overlapping local windows with window size $M \times M$. For each window $i$, we flatten and transpose its feature maps to obtain $\mathbf{X}^i \in \mathbb{R}^{M^2 \times C}$. Let $\mathbf{X}= \left\{ \mathbf{X}^1, \mathbf{X}^2,...,\mathbf{X}^N  \right\}, N=HW/M^2$. Next, we use projection matrices $\mathbf{W}^Q, \mathbf{W}^K, \mathbf{W}^V \in \mathbb{R}^{C \times 1}$ for the queries, keys, and values, respectively. Then, we calculate the query, key, and value $\mathbf{Q}^i=\mathbf{X}^i\mathbf{W^Q}$, $\mathbf{K}^i=\mathbf{X}^i\mathbf{W^K}$, and $\mathbf{V}^i=\mathbf{X}^i\mathbf{W^V}$. The W-SA mechanism can be defined as follows:
    
    \begin{equation}
    \begin{aligned}
    \mathbf{W \textendash SA}(\mathbf{Q}^i, \mathbf{K}^i, \mathbf{V}^i) = \mathbf{Softmax}(\mathbf{Q}^i\mathbf{K}^{i\mathbf{T}}+\mathbf{B}) \times \mathbf{V}^i,
    \end{aligned}
    \end{equation}
    where the learnable parameter $\mathbf{B}$ is the relative position encoding bias. We apply the W-SA mechanism to each separated window within $\mathbf{X}$ and then merge them back to the original shape.
    
    \paragraph{Computational Complexity}
    The computational complexities of S-SA and W-SA  are described as follows:
    \begin{equation}
    \begin{aligned}
    O(\mathbf{S \textendash SA}) = \frac{HWC^2}{N}\\
    O(\mathbf{W \textendash SA}) = M^2 HWC.
    \end{aligned}
    \end{equation} 
    The computational complexity of S-SA and W-SA is linear for the spatial dimension HW. However, the computational complexity of W-SA grows linearly, while S-SA grows quadratically for channel dimension C.

    Through the design of the SLSST low-rank model architecture, we can avoid performing self-attention on high-dimensional feature maps. Instead, perform self-attention computation on the basis component with lower spatial dimension and abundance component with lower channel dimension, significantly reducing the computational cost.
    
    \begin{figure}[t]
		\centerline{\includegraphics[width=0.49\textwidth]{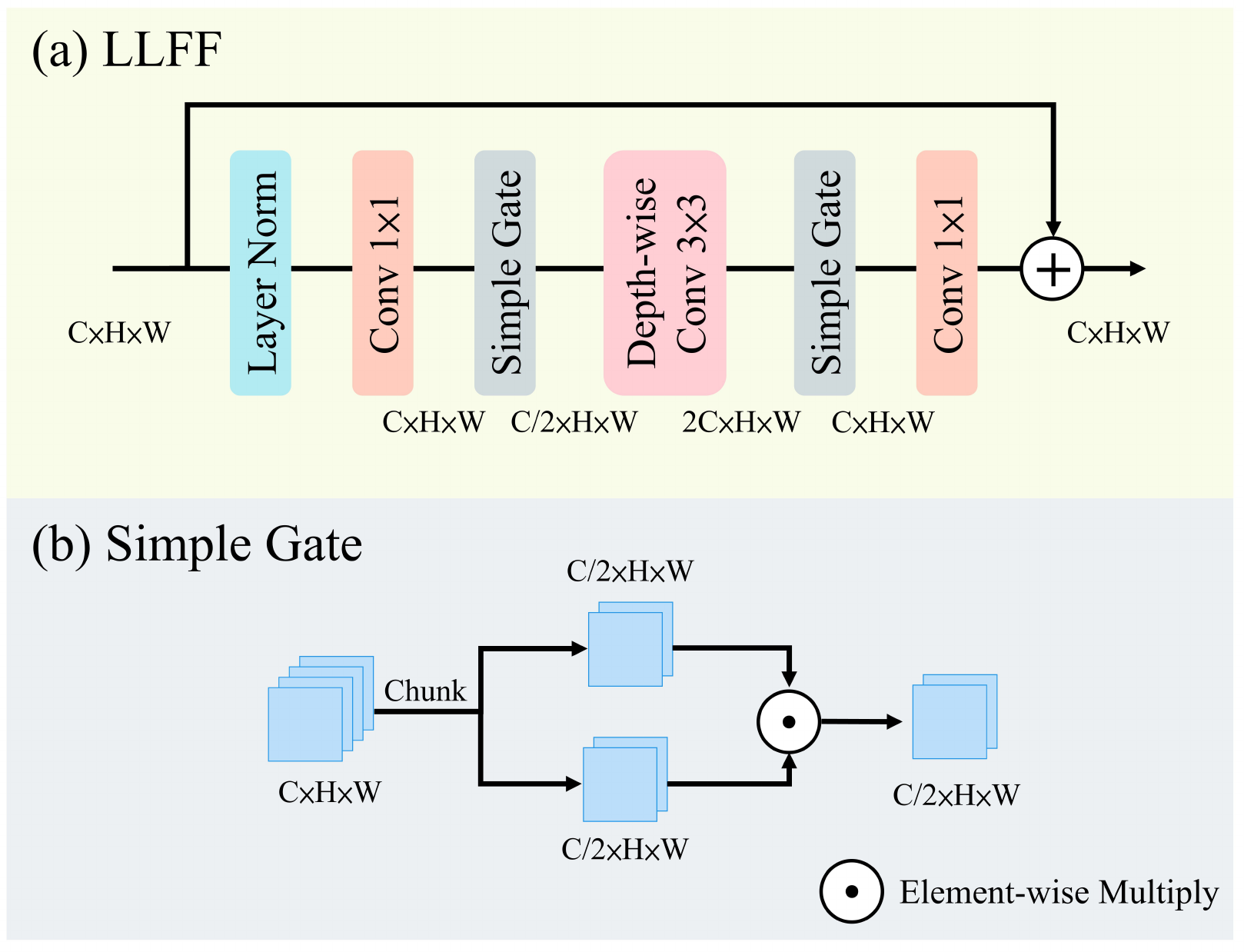}}
		\caption{{Illustration of (a) lightweight locally-enhanced feed-
        forward network and (b) Simple Gate.}
		}
		\label{fig:llff}
	\end{figure}
 
	\subsubsection{Lightweight Locally-enhanced Feed-Forward Network (LLFF)}
    The Feed-Forward Network (FFN) is one of the primary modules in the Transformer. As previously mentioned, S-SA and W-SA are mainly used to acquire global long-range information. Enhancing local context information is also crucial for HSI restoration. Thus, we proposed the LLFF, which can achieve this goal with lightweight computational cost. 

    As shown in Fig. \ref{fig:llff}(a), LLFF comprises 1 × 1 convolution, Simple Gate \cite{chen2022simple}, and depth-wise convolution. Unlike vanilla FFNs that use Gaussian error linear units (GELU) \cite{hendrycks2016gaussian} to provide non-linearity, we use Simple Gate to replace the computationally expensive GELU. The Simple Gate operation is easy to implement as shown in Fig. \ref{fig:llff}(b). To use it, we split the input feature maps $\mathbf{X}_{in} \in \mathbb{R}^{C \times H \times W}$ into two parts along the channel dimension, $\mathbf{X}, \mathbf{Y} \in \mathbb{R}^{(C/2) \times H \times W}$, and multiply them element-wise as follows:
    \begin{equation}
    \begin{aligned}
    \mathbf{Simple \; Gate}(\mathbf{X},\mathbf{Y}) = \mathbf{X} \odot \mathbf{Y},
    \end{aligned}
    \end{equation}
    where $\odot$ represents element-wise multiplication.

    LLFF uses 1 × 1 convolution for projection, Simple Gate for non-linearity, and depth-wise convolution for enhancing local information, as well as to compensate for the reduced feature maps dimension due to the operation of the Simple Gate. By using LLFF, we can enhance feature locality without significantly increasing the computational cost.


    \begin{table*}[t]
    \centering
    \caption{Quantitative comparisons of various HSI  denoising methods on Gaussian noise intensity $\sigma=$30, 50, 70, and blind (random range from 30-70).}
    \label{tab:denoise}
    \renewcommand\arraystretch{1.2}{
    \setlength{\tabcolsep}{2.0mm}{
    \begin{tabular}{cccccccccccc}
    \hline
    $\sigma$ &
      Metric &
      Noisy &
      LRTDTV \cite{wang2017hyperspectral}&
      FastHyDe \cite{zhuang2018fast}&
      NGmeet \cite{he2019non}&
      DHP \cite{sidorov2019deep}&
      T3SC \cite{bodrito2021trainable}&
      AODN \cite{kan2021attention}&
      RCTV \cite{peng2022fast}&
      SST \cite{li2022spatial}&
      Proposed \\ \hline
    30 &
      \begin{tabular}[c]{@{}c@{}}MPSNR $\uparrow$ \\ MSSIM $\uparrow$ \\ SAM $\downarrow$ \end{tabular} &
      \begin{tabular}[c]{@{}c@{}}19.779\\ 0.137\\ 29.209\end{tabular} &
      \begin{tabular}[c]{@{}c@{}}37.236\\ 0.897\\ 3.932\end{tabular} &
      \begin{tabular}[c]{@{}c@{}}27.924\\ 0.646\\ 9.539\end{tabular} &
      \begin{tabular}[c]{@{}c@{}}29.790\\ 0.728\\ 7.844\end{tabular} &
      \begin{tabular}[c]{@{}c@{}}32.614\\ 0.803\\ 6.616\end{tabular} &
      \begin{tabular}[c]{@{}c@{}}41.324\\ \bf{0.946}\\ 3.020\end{tabular} &
      \begin{tabular}[c]{@{}c@{}}37.312\\ 0.886\\ 4.705\end{tabular} &
      \begin{tabular}[c]{@{}c@{}}38.938\\ 0.918\\ 3.394\end{tabular} &
      \begin{tabular}[c]{@{}c@{}}39.736\\ 0.931\\ 3.578\end{tabular} &
      \begin{tabular}[c]{@{}c@{}}\bf{41.947}\\ 0.944\\ \bf{2.844}\end{tabular} \\
    \hline
    50 &
      \begin{tabular}[c]{@{}c@{}}MPSNR $\uparrow$\\ MSSIM $\uparrow$\\  SAM $\downarrow$\end{tabular} &
      \begin{tabular}[c]{@{}c@{}}15.831\\ 0.061\\ 38.334\end{tabular} &
      \begin{tabular}[c]{@{}c@{}}35.202\\ 0.856\\ 4.967\end{tabular} &
      \begin{tabular}[c]{@{}c@{}}24.692\\ 0.4554\\ 14.330\end{tabular} &
      \begin{tabular}[c]{@{}c@{}}26.624\\ 0.671\\ 11.131\end{tabular} &
      \begin{tabular}[c]{@{}c@{}}28.615\\ 0.774\\ 10.239\end{tabular} &
      \begin{tabular}[c]{@{}c@{}}39.130\\ \bf{0.920}\\ 3.517\end{tabular} &
      \begin{tabular}[c]{@{}c@{}}35.183\\ 0.838\\ 5.922\end{tabular} &
      \begin{tabular}[c]{@{}c@{}}36.500\\ 0.877\\ 4.671\end{tabular} &
      \begin{tabular}[c]{@{}c@{}}38.052\\ 0.907\\ 3.877\end{tabular} &
      \begin{tabular}[c]{@{}c@{}}\bf{39.854}\\ 0.917\\ \bf{3.316}\end{tabular} \\
    \hline
    70 &
      \begin{tabular}[c]{@{}c@{}}MPSNR $\uparrow$\\ MSSIM $\uparrow$\\ SAM $\downarrow$\end{tabular} &
      \begin{tabular}[c]{@{}c@{}}13.284\\ 0.034\\ 43.753\end{tabular} &
      \begin{tabular}[c]{@{}c@{}}33.885\\ 0.826\\ 5.762\end{tabular} &
      \begin{tabular}[c]{@{}c@{}}21.868\\ 0.327\\ 17.907\end{tabular} &
      \begin{tabular}[c]{@{}c@{}}23.852\\ 0.623\\ 14.008\end{tabular} &
      \begin{tabular}[c]{@{}c@{}}25.091\\ 0.731\\ 13.391\end{tabular} &
      \begin{tabular}[c]{@{}c@{}}37.461\\ 0.894\\ 4.020\end{tabular} &
      \begin{tabular}[c]{@{}c@{}}33.725\\ 0.773\\ 6.404\end{tabular} &
      \begin{tabular}[c]{@{}c@{}}34.812\\ 0.841\\ 5.910\end{tabular} &
      \begin{tabular}[c]{@{}c@{}}37.922\\ 0.892\\ 4.008\end{tabular} &
      \begin{tabular}[c]{@{}c@{}}\bf{38.703}\\ \bf{0.900}\\ \bf{3.605}\end{tabular} \\
    \hline
    Blind &
      \begin{tabular}[c]{@{}c@{}}MPSNR $\uparrow$\\ MSSIM $\uparrow$\\ SAM $\downarrow$\end{tabular} &
      \begin{tabular}[c]{@{}c@{}}16.377\\ 0.073\\ 37.026\end{tabular} &
      \begin{tabular}[c]{@{}c@{}}35.498\\ 0.863\\ 4.847\end{tabular} &
      \begin{tabular}[c]{@{}c@{}}24.970\\ 0.480\\ 13.786\end{tabular} &
      \begin{tabular}[c]{@{}c@{}}26.887\\ 0.675\\ 10.836\end{tabular} &
      \begin{tabular}[c]{@{}c@{}}28.656\\ 0.750\\ 10.016\end{tabular} &
      \begin{tabular}[c]{@{}c@{}}38.402\\ 0.912\\ \bf{3.498}\end{tabular} &
      \begin{tabular}[c]{@{}c@{}}34.103\\ 0.782\\ 5.958\end{tabular} &
      \begin{tabular}[c]{@{}c@{}}36.796\\ 0.881\\ 4.573\end{tabular} &
      \begin{tabular}[c]{@{}c@{}}37.991\\ 0.905\\ 4.631\end{tabular} &
      \begin{tabular}[c]{@{}c@{}}\bf{39.366}\\ \bf{0.915}\\ 3.642\end{tabular} \\ \hline
    \end{tabular}}}
    \end{table*}

    \begin{figure*}[t]
		\centerline{\includegraphics[width=1.03\textwidth]{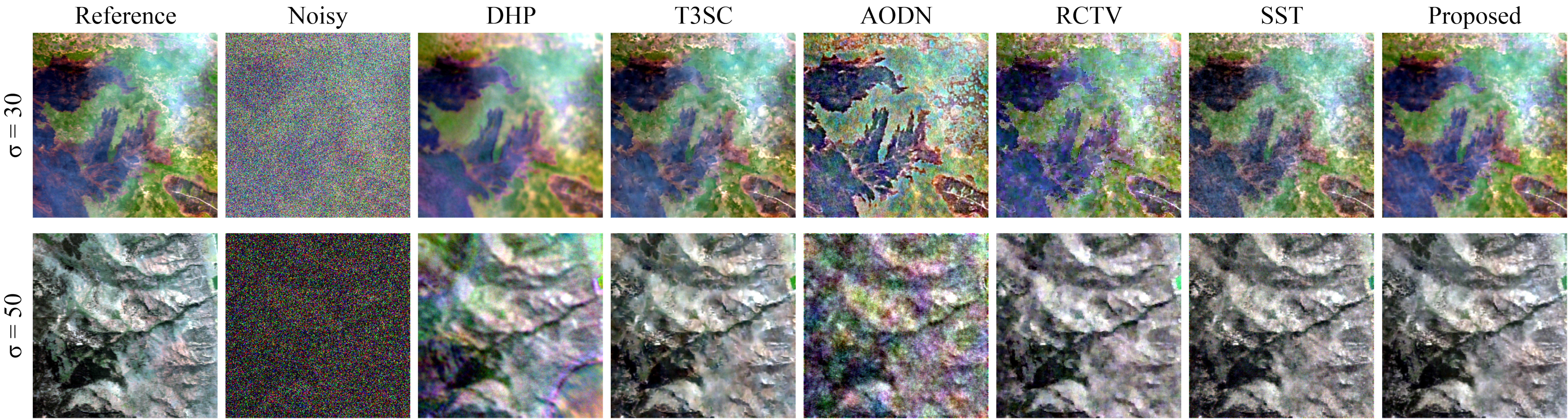}}
		\caption{{Visual comparisons of various HSI denoising methods, with $\sigma=$30 and 50 on HSIs acquired over Little Bear Ray, USA and Harney Basin, USA, respectively.}}
		\label{fig:denoise}
	\end{figure*}

    \begin{figure*}[t]
		\centerline{\includegraphics[width=1.03\textwidth]{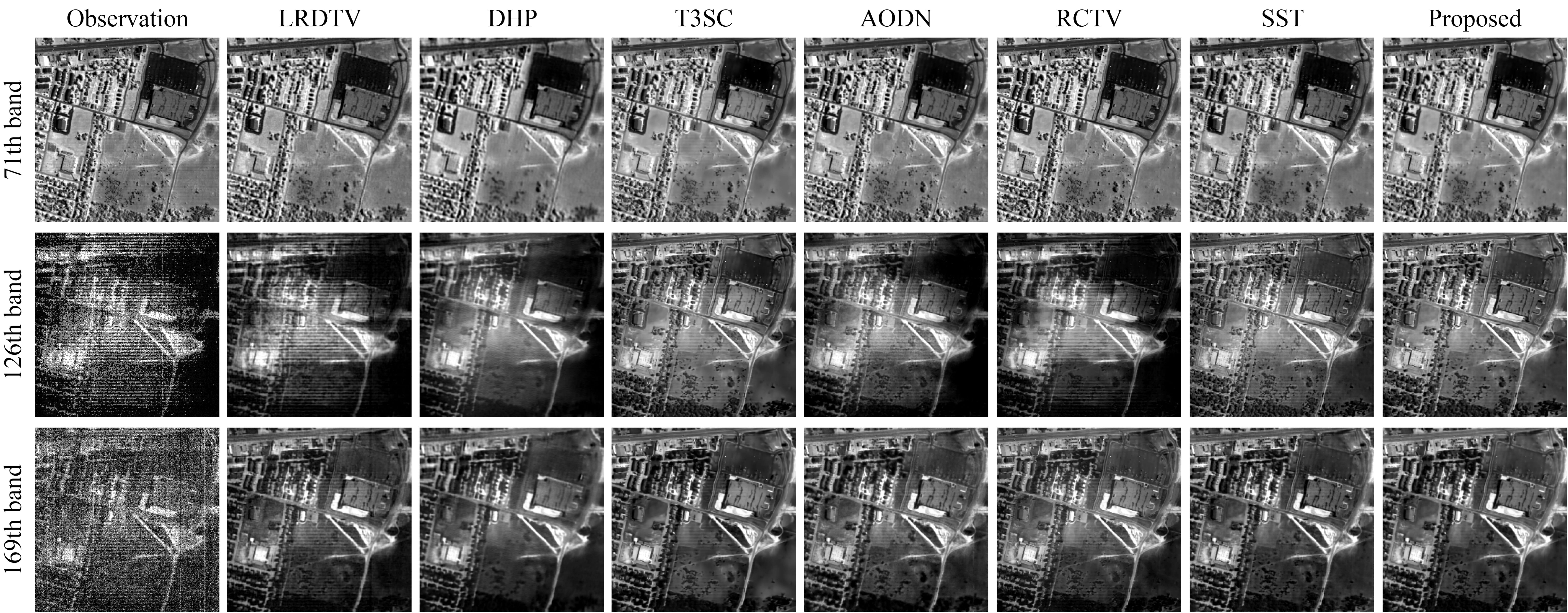}}
		\caption{{Visual comparisons of various HSI denoising methods on real HSI Urban dataset.}}
		\label{fig:denoise_real}
	\end{figure*}
 
    \section{Experimental Results}\label{sec:experiment}
 
	
     
     In this section, we will first introduce the remote sensing HSI dataset we used in Section \ref{sec:datasets} and describe our experimental setup in Section \ref{sec:setting}. Then, we will verify the effectiveness of the proposed Hyper-Restormer on various HSI restoration tasks on multiple HSI datasets, HSI denoising in Section \ref{sec:denoise}, HSI inpainting in Section \ref{sec:inpaint}, and HSI super-resolution in Section \ref{sec:sr}. Finally, we will conduct ablation studies on the proposed components in Section \ref{sec:ablation}, and the computational time required by the model will be presented in Section \ref{sec:time}.

    \subsection{Datasets}\label{sec:datasets}

    \subsubsection{Simulation Data}
     The remote sensing HSIs used for the simulation experiments were acquired from the Airborne Visible/Infrared Imaging Spectrometer (AVIRIS) sensor \cite{AVIRISrealdata}, which consisted of 224 spectral bands. After removing spectral bands 1-10, 104-116, 152-170, and 215-224 \cite{lin2021admm}, the HSI had a spatial size of 256 x 256 pixels and 172 spectral bands. The simulation data include various terrains such as cities, mountains, vegetation, and lakes in the USA and Canada, acquired from 2008 to 2018.

    In the simulation experiments, a total of 875 HSIs were used. We randomly select 800 HSIs for deep learning training and 75 HSIs for testing. Unlike many remote sensing HSI restoration methods that partition a HSI into numerous small patches for training and testing, we utilize large-sized images and diverse terrain structures to avoid overfitting to a single type of terrain for more accurate evaluation.

    The HSIs are subjected to Gaussian noise, random stripe, and downsampling for training and testing according to different experimental categories.
    
    \subsubsection{Real Data}
    As for the real data experiments, we selected one dataset for each of the three hyperspectral restoration tasks to evaluate our method.
    
    For the denoising task, we used the Urban dataset \cite{Urban}, which is commonly used for HSI denoising tasks and contains unknown noise. The Urban dataset was captured by the HYDICE sensor, which has 210 spectral bands. To prepare the data, we discarded spectral bands 1-7, 67-77, 122-128, and 166-178, and cropped the central portion of the dataset, resulting in HSI data with a spatial size of 256 x 256 pixels and 172 spectral bands.

    In the inpainting task, we employed commonly studied HSI inpainting data from Bhilwara, India \cite{Bhilwara} that was captured by the Hyperion sensor onboard NASA's Earth Observing-1 (EO-1) satellite \cite{Hyperion}, which has 242 spectral bands. After removing spectral bands 1–7, 61–77, 122–128, 166–178, and 217–242, the data had a spatial size of 256 x 256 pixels and 172 spectral bands.

    Regarding the super-resolution task, we utilized the Washington DC Mall dataset \cite{WDC}, a frequently used HSI dataset. It was captured by the HYDICE sensor and consisted of 191 spectral bands.
    After excluding spectral bands 173-191, we selected a section of the HSI and cropped it to a size of 32 x 32 pixels. Consequently, the resulting HSI data had a spatial size of 32 x 32 pixels and 172 spectral bands.
    
    \subsection{Experimental Setting}\label{sec:setting} 
    In Hyper-Restormer, we adopt a multi-stage restoration strategy by cascading multiple SLSSTs. The number of SLSSTs $N_S$ is set to 4. The window size for the Window-based Self-Attention in the model is set to 8. Additionally, the embedding dimension of SLSST $E$ is set to 172.
    
    For each experiment, we employed the same model and hyperparameters. The training process lasted for 300 epochs with a batch size of 8. We used AdamW \cite{loshchilov2017decoupled} with $\beta_1 = 0.9$ and $\beta_2 = 0.999$ as the optimizer with an initial learning rate of $3 \times 10^{-4}$ and applied the cosine annealing strategy to gradually reduce the learning rate from $3 \times 10^{-4}$ to $1 \times 10^{-6}$. We train our model using the L1 loss.
    
    During the deep learning training phase, we used cloud computing containers in Python 3.8.10 environment equipped with NVIDIA Tesla V100 32GB GPU and Intel Xeon Gold 6154 CPU (3.00-GHz speed and 60-GB RAM). In the testing phase, all experiments were conducted on a desktop computer equipped with NVIDIA RTX-3090 24GB GPU and Intel Core-i9-10900K CPU (3.70-GHz speed and 64-GB RAM). The computational environment for deep learning was implemented on Python 3.7.11, while all other methods were executed on Mathworks Matlab R2021a.

   We evaluated the experimental results with commonly used quantitative metrics mean peak signal-to-noise ratio (MPSNR) \cite{liu2017universal}, mean structural similarity (MSSIM) \cite{yuan2018hyperspectral}, and spectral angle mapper (SAM) \cite{kruse1993spectral}. Higher values of MPSNR and MSSIM indicate better performance, while a lower value of SAM indicates better performance.

	\subsection{Remote Sensing HSI Denoising}\label{sec:denoise}
    For the denoising simulation experiment, we added Gaussian noise with four different noise levels to the input HSI, including $\sigma=$ 30, 50, 70, and blind (random range from 30-70).
    Table \ref{tab:denoise} reports the results of the HSI denoising task. We compared Hyper-Restormer with eight state-of-the-art HSI denoising methods, including optimized-based methods LRDTV, FastHyDe, NGmeet, RCTV, and deep learning-based methods DHP, T3SC, AODN, and SST.
    Our proposed method outperforms other methods in most quantitative metrics, with only a slight lag on a few metrics to T3SC, fully demonstrating the superior denoising performance of Hyper-Restormer.
    It is worth noting that while SST is capable of effectively using non-local spatial self-attention and global spectral self-attention mechanisms to improve restoration performance, its high computational complexity makes it impractical to use the full size of the HSI for training, thereby limiting its ability to utilize the complete information of the HSI.
   
    The denoising results for $\sigma=$ 30 and 50 are shown in Figure \ref{fig:denoise}, and it can be observed that our method has excellent visual performance. T3SC also performs well, but there are some blurring artifacts in the details. In contrast, SST images have a large amount of fine noise present. The denoising results on a real HSI using the Urban dataset are also shown in Figure \ref{fig:denoise_real}. We applied a noise level of $\sigma=30$ pre-trained model and parameter setting for denoising. Our method achieves clean removal of noise except deadline noise that was not present in the training data, while the results obtained by T3SC are somewhat over-smoothed, leading to a loss of some details.



    \begin{table*}[t]
    \centering
    \caption{Quantitative comparisons of various HSI inpainting methods on corrupted data with random stripe patterns and missing bands.}
    \label{tab:inpaint}
    \renewcommand\arraystretch{1.2}{
    \setlength{\tabcolsep}{4.2mm}{
    \begin{tabular}{cccccccccc}
    \hline
    Metric &
      3D-PDE \cite{d2008inpainting}&
      UBD \cite{cerra2014unmixing}&
      LLRSSTV \cite{he2018hyperspectral}&
      FastHyIn \cite{zhuang2018fast}&
      DHP \cite{sidorov2019deep}&
      ADMM-ADAM \cite{lin2021admm}&
      Proposed \\ \hline
    \begin{tabular}[c]{@{}c@{}}MPSNR $\uparrow$\\ MSSIM $\uparrow$\\ SAM $\downarrow$ \end{tabular} &
      \begin{tabular}[c]{@{}c@{}}44.723\\ 0.964\\ 6.621\end{tabular} &
      \begin{tabular}[c]{@{}c@{}}38.297\\ 0.937\\ 8.464\end{tabular} &
      \begin{tabular}[c]{@{}c@{}}38.872\\ 0.9365\\ 7.131\end{tabular} &
      \begin{tabular}[c]{@{}c@{}}52.398\\ 0.987\\ 5.127\end{tabular} &
      \begin{tabular}[c]{@{}c@{}}45.987\\ 0.960\\ 8.235\end{tabular} &
      \begin{tabular}[c]{@{}c@{}}51.365\\ 0.990\\ 1.246\end{tabular} &
      \begin{tabular}[c]{@{}c@{}}\bf{52.631}\\ \bf{0.993}\\ \bf{1.076}\end{tabular} \\ \hline
    \end{tabular}}}
    \end{table*}

    \begin{figure*}[t]
		\centerline{\includegraphics[width=1.03\textwidth]{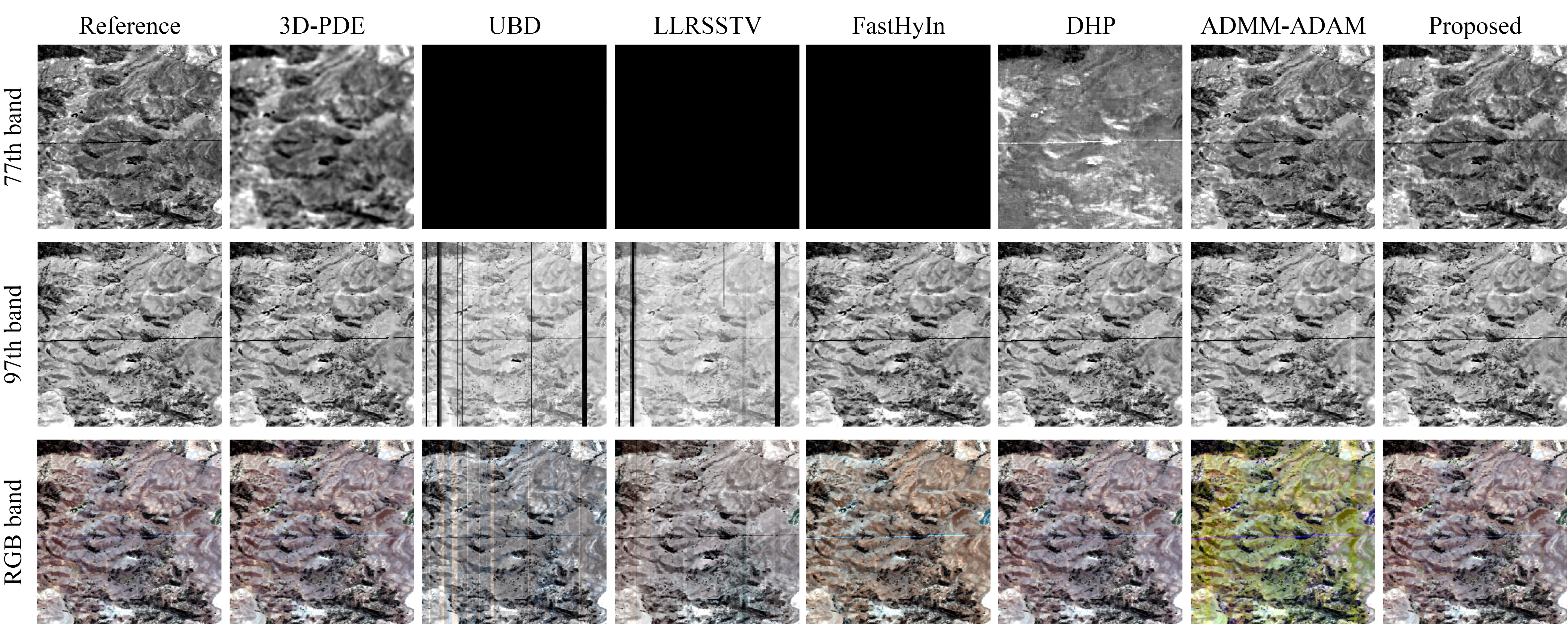}}
		\caption{{Visual comparisons of various HSI inpainting methods, on HSI acquired over Lassen National Forest, USA. The missing stripe patterns are visualized in Fig. \ref{fig:mask}. }}
		\label{fig:inpaint}
	\end{figure*}

    \begin{figure}[h!]
		\centerline{\includegraphics[width=0.49\textwidth]{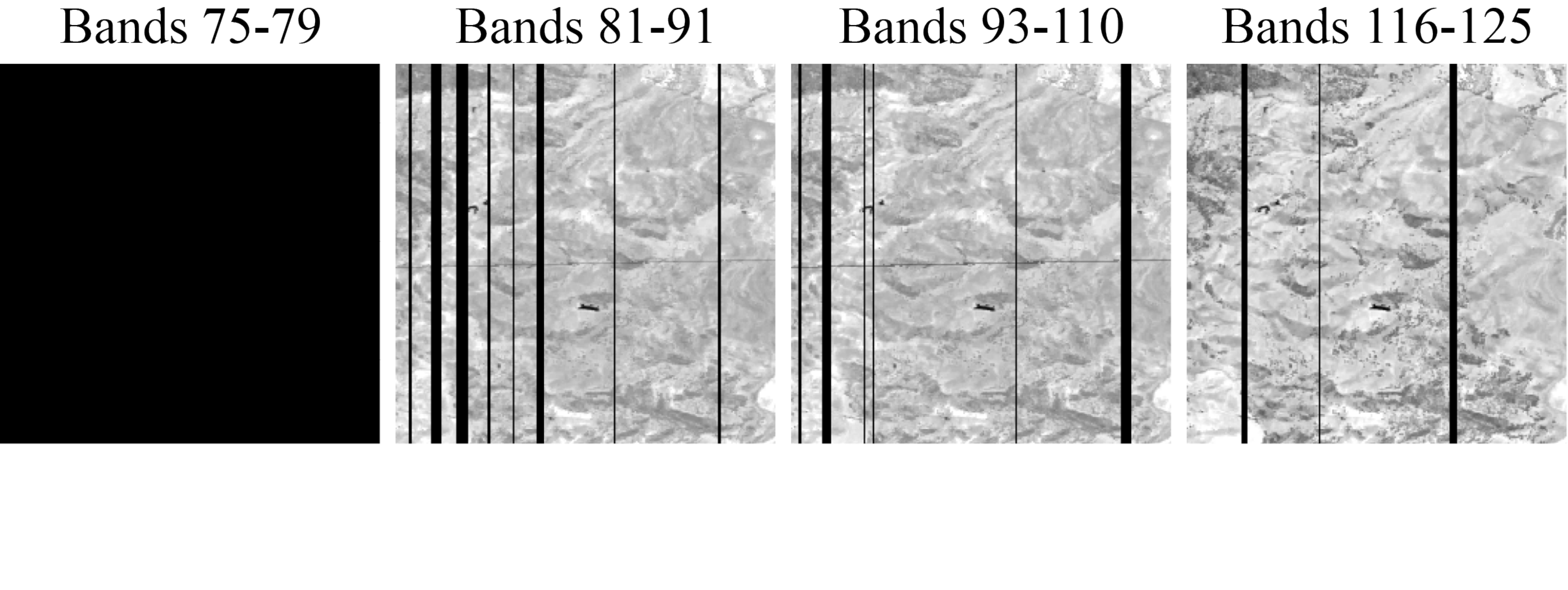}}
		\caption{The stripe patterns of the HSI acquired over Lassen National Forest, USA.}
		\label{fig:mask}
	\end{figure}

	\begin{figure*}[t]
		\centerline{\includegraphics[width=1.03\textwidth]{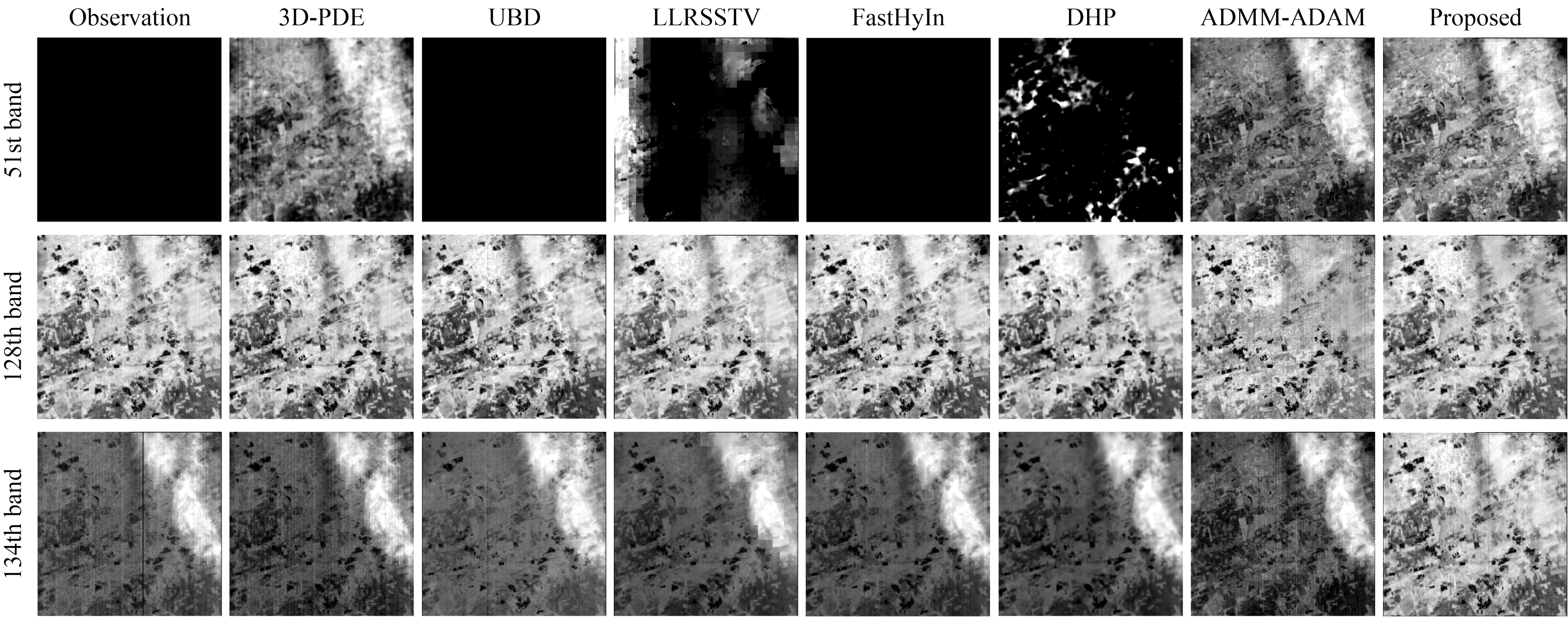}}
		\caption{{Visual comparisons of various HSI inpainting methods on real HSI Bhilwara, India from NASA's Hyperion.}}
		\label{fig:inpaint_real}
	\end{figure*}

	\subsection{Remote Sensing HSI Inpainting}\label{sec:inpaint}
    In the inpainting simulation experiments, we simulated HSI damage by creating random striped patterns with continuous bands in the input HSI. The width, position, and number of stripes generated were also random. We introduced randomly continuous missing bands, which created a highly challenging scenario for inpainting algorithms.
    We present the inpainting results in Table \ref{tab:inpaint}. Hyper-Restormer is compared with six state-of-the-art HSI inpainting methods, including 3D-PDE, UBD, LLRSSTV, FastHyIn, DHP, and ADMM-ADAM.
    
    Our method outperforms all inpainting methods in three quantitative metrics, while FastHyIn and ADMM-ADAM also achieve excellent performance. However, in the inpainting task, one can achieve high performance in quantitative metrics by generating very similar results to the input damaged image. Therefore, it is necessary to analyze the results in combination with the visualized results.
    
    We show the visual inpainting results in Fig. \ref{fig:inpaint}, and we can observe that 3D-PDE has good visual performance, but there is blurring in severely damaged areas. Although FastHyIn has good metric performance, its biggest weakness, like some of the other methods, is that it cannot reconstruct completely missing bands, and also exist color deviations in the RGB band. DHP has good visual performance but cannot tackle completely missing bands effectively. Only ADMM-ADAM and our method can effectively reconstruct the lost spectral information. However, ADMM-ADAM still shows more visible damage traces in some bands and has color deviation issues too. 

    The results on real data from Bhilwara, India, are shown in Figure \ref{fig:inpaint_real}, and for the completely missing information in the 51st band, only ADMM-ADAM and our method can effectively recover it. However, on the relatively clean 128th band, ADMM-ADAM shows traces of stripe damage, resulting in worse visual effects than other methods. On the 134th band, our method not only reconstructed most of the missing areas but also solved the problem of low brightness and cloud obstruction encountered during shooting. These results prove that our method indeed achieves state-of-the-art inpainting performance.

	\begin{table*}[t]
    \centering
    \caption{Quantitative comparisons of various HSI super-resolution methods on 4x and 8x spatial super-resolution.}
    \label{tab:sr}
    \renewcommand\arraystretch{1.2}{
    \setlength{\tabcolsep}{2.2mm}{
    \begin{tabular}{cccccccccc}
    \hline
    Scale &
      Metric &
      {Bicubic} &
      {3D-FCNN \cite{mei2017hyperspectral}} &
      {GDRRN \cite{li2018single}} &
      {DHP \cite{sidorov2019deep}} &
      {3D-GAN \cite{li2020hyperspectral}} &
      {SSPSR \cite{jiang2020learning}} &
      {ADMM-ADAM SR \cite{lin2022single}} &
      {Proposed} \\ \hline
    4 &
      \begin{tabular}[c]{@{}c@{}}MPSNR $\uparrow$\\ MSSIM $\uparrow$\\ SAM $\downarrow$\end{tabular} &
      \begin{tabular}[c]{@{}c@{}}36.620\\ 0.757\\ 3.214\end{tabular} &
      \begin{tabular}[c]{@{}c@{}}36.341\\ 0.805\\ 3.434\end{tabular} &
      \begin{tabular}[c]{@{}c@{}}36.724\\ 0.808\\ 3.159\end{tabular} &
      \begin{tabular}[c]{@{}c@{}}33.113\\ 0.649\\ 5.351\end{tabular} &
      \begin{tabular}[c]{@{}c@{}}35.252\\ 0.782\\ 3.686\end{tabular} &
      \begin{tabular}[c]{@{}c@{}}36.995\\ 0.780\\ 3.206\end{tabular} &
      \begin{tabular}[c]{@{}c@{}}36.803\\ 0.743\\ 3.968\end{tabular} &
      \begin{tabular}[c]{@{}c@{}}\bf{37.600}\\ \bf{0.835}\\ \bf{2.873}\end{tabular} \\
    \hline
    8 &
      \begin{tabular}[c]{@{}c@{}}MPSNR $\uparrow$\\ MSSIM $\uparrow$\\  SAM $\downarrow$\end{tabular} &
      \begin{tabular}[c]{@{}c@{}}33.615\\0.690\\4.565\end{tabular} &
      \begin{tabular}[c]{@{}c@{}}33.600\\ 0.707\\ 4.670\end{tabular} &
      \begin{tabular}[c]{@{}c@{}}33.937\\ 0.710\\ 4.424\end{tabular} &
      \begin{tabular}[c]{@{}c@{}}30.550\\ 0.596\\ 7.253\end{tabular} &
      \begin{tabular}[c]{@{}c@{}}33.493\\ 0.707\\ 4.668\end{tabular} &
      \begin{tabular}[c]{@{}c@{}}34.108\\ 0.689\\ 4.501\end{tabular} &
      \begin{tabular}[c]{@{}c@{}}34.581\\ 0.681\\ 4.726\end{tabular} &
      \begin{tabular}[c]{@{}c@{}}\bf{36.164}\\ \bf{0.775}\\ \bf{3.535}\end{tabular} \\ \hline  
    \end{tabular}}}
    \end{table*}
	
	\begin{figure}[h!]
    		\centerline{\includegraphics[width=0.49\textwidth]{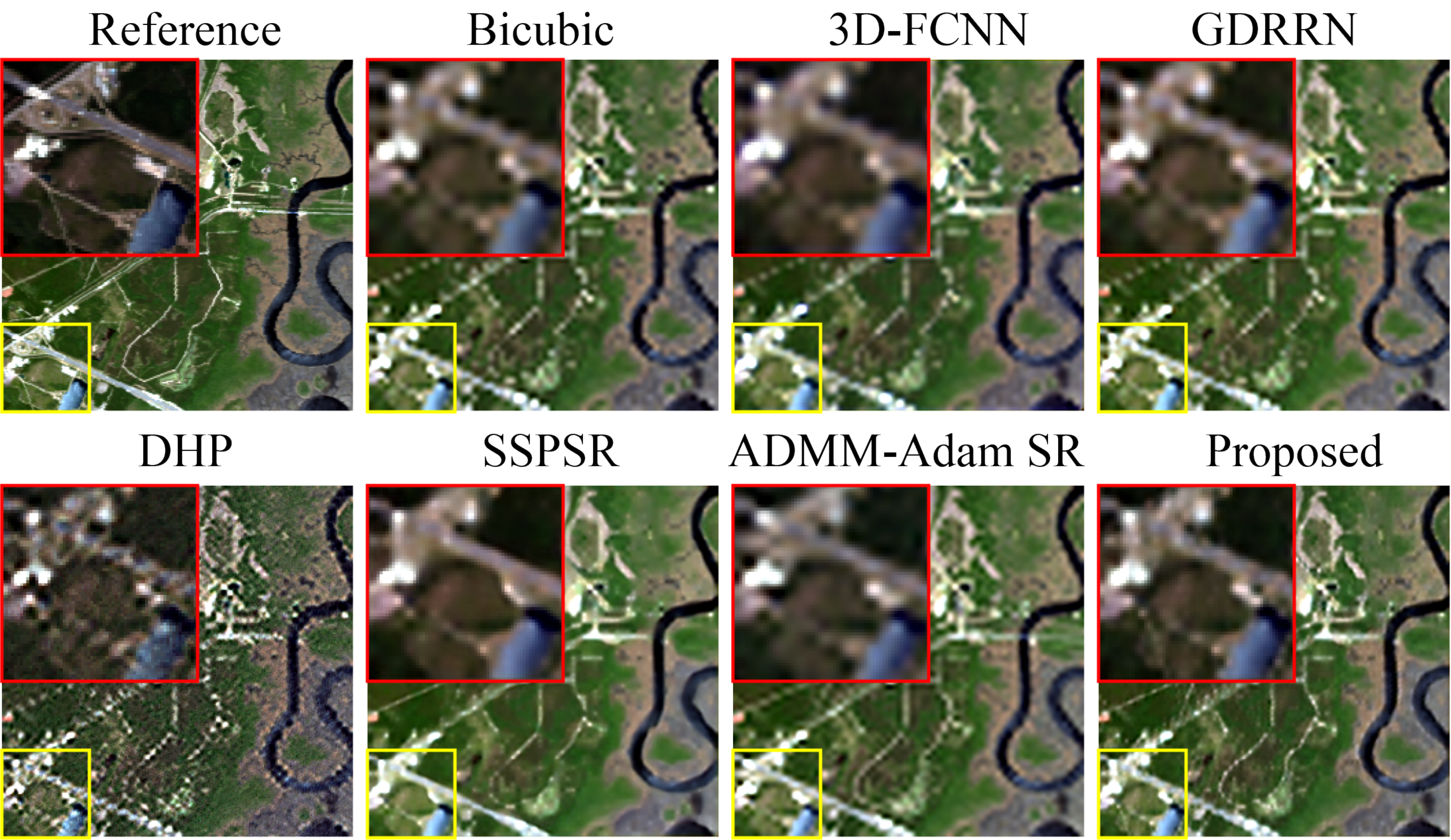}}
    		\caption{
                Visual comparisons of various HSI super-resolution methods, on HSI acquired over Osceola Natural Area, USA with 4x spatial super-resolution.}
    		\label{fig:sr}
    	\end{figure}
     
     \begin{figure}[h!]
    		\centerline{\includegraphics[width=0.49\textwidth]{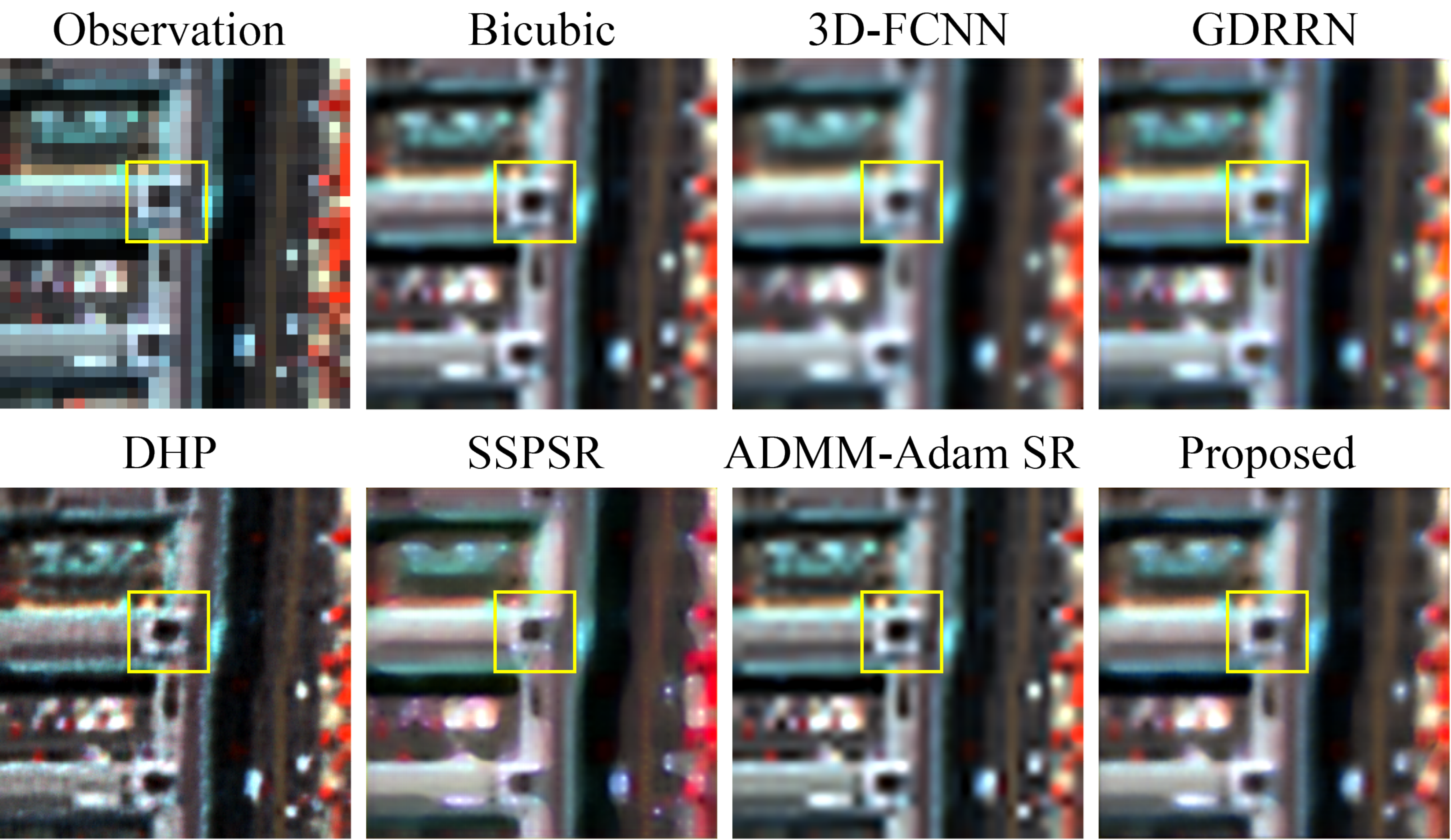}}
    		\caption{
                Visual comparisons of various HSI super-resolution methods, on real HSI Washington DC Mall dataset with 8x spatial super-resolution.}
    		\label{fig:sr_real}
    	\end{figure}

	\subsection{Remote Sensing HSI Super-Resolution}\label{sec:sr}
    In the simulation super-resolution experiments, we first downsampled the original spatial resolutions of the 256x256 HSI to 64x64 and 32x32 for low spatial resolution input. The input is then applied to the super-resolution method to obtain the original 256x256 spatial-size HSI. We compared our method with six state-of-the-art deep learning-based super-resolution methods: 3D-FCNN, GDRRN, DHP, 3D-GAN, SSPSR, and ADMM-Adam SR. ADMM-Adam SR further combines deep learning with convex optimization.
    For 3D-FCNN, GDRRN, and our proposed method, the input HSI is first upsampled to the same size as the output image by bicubic interpolation before input to the model. 
    
    Table \ref{tab:sr} shows the results of HSI super-resolution. 3D-FCNN, DHP, and 3D-GAN seem not able to obtain better results than bicubic interpolation. We speculate that the feature extraction capability of the 3D-FCNN model is not powerful enough to achieve better performance. In addition, we use a lot of testing data, so it was not possible to customize the optimal number of training iterations for each HSI in DHP experiment. It decreased the quality when the generated results deviated from the optimal number of iterations. The computational cost of spatial-spectral constraint loss in 3D-GAN is too high, resulting in its removal during training, which leads to poorer results.
    Our proposed Hyper-Restormer significantly outperforms other methods in all quantitative metrics, especially at 8x scale, demonstrating its effective performance.
    
    Figure \ref{fig:sr} shows the results of super-resolution methods at 4x scale, where Hyper-Restormer outperforms other methods in terms of visual quality, clearly restoring more detailed information in the zoom-in area. For real HSI experiments, we tested it on the Washington DC Mall dataset by selecting a 32x32 spatial size area from the original HSI. We then applied pre-trained models for 8x scale super-resolution, and the results are shown in Figure \ref{fig:sr_real}. It seems that only SSPSR and our method can restore more details of the roof area, but SSPSR's result contains many grid-like artifacts and slight spatial distortions. The rest of the methods either have some noise or are too smooth to lose details. Our method achieves a better balance in terms of visual performance.

    \begin{table}[t]
		\centering 
        \caption{Effect of Spectral Attention (Spe-A) Block and Spatial Attention (Spa-A) Block.}\label{tab:attention}
        \renewcommand\arraystretch{1.1}{
		\begin{tabular}{c|c|c|c|c}
            \hline
             Spe-A Block &Spa-A Block &MPSNR $\uparrow$ &MSSIM $\uparrow$ &SAM $\downarrow$\\
            \hline
                 {-}&{-}&{38.549}&{0.897}&{3.666} \\
             \CheckmarkBold &{-}&{38.563}&{0.895}&{3.692} \\
          {-}&\CheckmarkBold&{38.504}&{0.894}&{3.724} \\	
            \CheckmarkBold&\CheckmarkBold&\bf{38.703}&\bf{0.900}&\bf{3.605}\\	
            \hline
		\end{tabular}}
	\end{table}

    \begin{table}[t]
		\centering 
        \caption{Effect of Spectral Attention Block and Spatial Attention Block arrangement.}\label{tab:arrange}
        \renewcommand\arraystretch{1.1}{
		\begin{tabular}{c|c|c|c}
            \hline
             Arrangement &MSPNR $\uparrow$ &MSSIM $\uparrow$ &SAM $\downarrow$  \\
            \hline
                 Spatial-Spectral&{38.485}&{0.895}&{3.684} \\
             Spectral-Spatial&{38.560}&{0.894}&{3.718} \\
          Parallel&\bf{38.703}&\bf{0.900}&\bf{3.605}\\	
            \hline
		\end{tabular}}
	\end{table}

    \begin{table}[t]
		\centering 
        \caption{Effect of Lightweight Locally-enhanced Feed-forward Network (LLFF).}\label{tab:llff}
        \renewcommand\arraystretch{1.1}{
		\begin{tabular}{c|c|c|c}
            \hline
             LLFF &MPSNR $\uparrow$ &MSSIM $\uparrow$ &SAM $\downarrow$ \\
            \hline
                 {-}&{38.366}&{0.890}&{3.767} \\
             \CheckmarkBold&\bf{38.703}&\bf{0.900}&\bf{3.605} \\		
            \hline
		\end{tabular}}
	\end{table}

    \begin{table}[t]
		\centering 
        \caption{Effect of stage number.}\label{tab:stage}
        \renewcommand\arraystretch{1.1}{
		\begin{tabular}{c|c|c|c}
            \hline
             Stage Number &MPSNR $\uparrow$ &MSSIM $\uparrow$ &SAM $\downarrow$ \\
            \hline
                 1&{37.662}&{0.877}&{4.119} \\
             2&{38.242}&{0.891}&{3.825} \\
          3&{38.324}&{0.891}&{3.798} \\	
            4&\bf{38.703}&\bf{0.900}&{3.605} \\	
            5&{38.698}&{0.897}&\bf{3.591} \\	
            \hline
		\end{tabular}}
	\end{table}
		
    \subsection{Ablation Study}\label{sec:ablation}
    We conducted ablation studies to validate the effectiveness of the proposed components. We tested them in the HSI denoising task with a Gaussian noise level of $\sigma$ = 70. Specifically, we experimented with the efficacy of the Spectral Attention Block and Spatial Attention Block, as well as the arrangement of these blocks. We also evaluated the effects of the lightweight
    locally-enhanced feed-forward network and stage number on the performance.
    
    \subsubsection{Spectral Attention Block and Spatial Attention Block}
    We compared the quantitative metrics before and after incorporating Spectral Attention Block and Spatial Attention Block. Firstly, we remove Spectral Attention Block and Spatial Attention Block in LSS Transformer Block. We then evaluated the results of adding a Spectral Attention Block or a Spatial Attention Block, and finally, we tested with both Spectral and Spatial Attention Blocks simultaneously. As shown in Table \ref{tab:attention}, adding one of the blocks alone did not result in significant improvement while adding both blocks together led to a noticeable improvement in the quantitative metrics with MPSNR increase of 0.154 dB.

	\subsubsection{Spectral and Spatial Attention Block Arrangement}
    After confirming the usefulness of the Spectral Attention Block and Spatial Attention Block for HSI restoration, we investigated whether the arrangement order of attention blocks would affect the final results. We conducted three experiments with different arrangements of the Spectral and Spatial Attention Blocks: Spatial-Spectral sequential (Spatial-Spectral), Spectral-Spatial sequential (Spectral-Spatial), and Spectral-Spatial parallel (Parallel) arrangements. As shown in Table \ref{tab:arrange}, the Spectral-Spatial parallel arrangement achieved the best restoration result, with MPSNR improvements of 0.218 dB and 0.143 dB compared to the Spatial-Spectral sequential and Spectral-Spatial sequential arrangements, respectively.


    \subsubsection{Lightweight Locally-enhanced Feed-Forward Network}
    Apart from the attention blocks, we proposed the lightweight locally-enhanced feed-forward network (LLFF), which uses lightweight operations to enhance local information. To demonstrate that LLFF can indeed improve restoration performance, we compared the results with and without LLFF and displayed them in Table \ref{tab:llff}. Adding LLFF resulted in a significant improvement of MPSNR 0.337 dB, demonstrating that LLFF is indeed useful.


    \subsubsection{Stage Number}
    We also studied the performance of our multi-stage restoration strategy with different numbers of stages. We tested the results for stage numbers ranging from 1 to 5. As shown in Table \ref{tab:stage}, the quantitative metrics improved as the stage number increased. The best performance was achieved when the stage number was 4, and the results were almost the same as when the stage number was further increased. Therefore, we used stage number $N_S=$4 in our paper.




    \subsection{Computational Time}\label{sec:time}
    We conducted experiments on the computation time required for various methods in real data HSI denoising, HSI inpainting, and HSI super-resolution tasks, and the results are presented in Table \ref{tab:time_denoise}, \ref{tab:time_inpaint} and \ref{tab:time_sr}. Hyper-Restormer achieved the fastest speed in denoising and inpainting tasks. In the super-resolution task, our method is comparable to other deep learning-based methods, demonstrating its practicality.

    \begin{table}[t]
		\centering
        \caption{Computational time of various HSI denoising methods on the real HSI Urban dataset.}\label{tab:time_denoise}
		\setlength{\tabcolsep}{0.05mm}{ 
		\begin{tabular}{cc|cc}
            \hline
            &Methods$~$&$~~~${Time (sec.)}&$~~~$
            \\
            \hline
            &LRTDTV{\cite{wang2017hyperspectral}}&$~~~${116.475}&$~~~$
            \\
            &FastHyDe{\cite{zhuang2018fast}}$~$&$~~~${8.067}&$~~~$
            \\	
            &NGmeet{\cite{he2019non}}&$~~~${60.773}&$~~~$
            \\
            &DHP{\cite{sidorov2019deep}}&$~~~${76.767}&$~~~$
            \\
            &T3SC{\cite{bodrito2021trainable}}&$~~~${1.375}&$~~~$
            \\
            &AODN{\cite{kan2021attention}}$~$&$~~~${7.309}&$~~~$
            \\	
            &RCTV{\cite{peng2022fast}}$~$&$~~~${10.565}&$~~~$
            \\
            &SST{\cite{li2022spatial}}$~$&$~~~${1.476}&$~~~$
            \\
            &Proposed{}$~$&$~~~$\bf{0.714}&$~~~$
            \\
            \hline
		\end{tabular}}
	\end{table}

    \begin{table}[t]
		\centering
        \caption{Computational time of various HSI inpainting methods on the real HSI Bhilwara, India.}\label{tab:time_inpaint}
		\setlength{\tabcolsep}{0.05mm}{ 
			\begin{tabular}{cc|cc}
				\hline
				&Methods$~$&$~~~${Time (sec.)}&$~~~$
				\\
				\hline
				&3D-PDE{\cite{d2008inpainting}}&$~~~${6.601}&$~~~$
				\\
				&UBD{\cite{cerra2014unmixing}}$~$&$~~~${19.023}&$~~~$
				\\	
				&LLRSSTV{\cite{he2018hyperspectral}}&$~~~${170.263}&$~~~$
				\\
				&FastHyIN{\cite{zhuang2018fast}}&$~~~${12.294}&$~~~$
				\\
				&DHP{\cite{sidorov2019deep}}&$~~~${306.955}&$~~~$
				\\
                &ADMM-ADAM{\cite{lin2021admm}}&$~~~${5.761}&$~~~$
                \\
                &Proposed{}$~$&$~~~$\bf{0.698}&$~~~$
                \\
				\hline
		\end{tabular}}
	\end{table}

    \begin{table}[t]
		\centering
        \caption{Computational time of various HSI super-resolution methods on the real HSI Washington DC Mall dataset.}\label{tab:time_sr}
		\setlength{\tabcolsep}{0.05mm}{ 
		\begin{tabular}{cc|cc}
            \hline
            &Methods&$~~~${Time (sec.)}&$~~~$
            \\
            \hline
            &Bicubic{}&$~~~$\bf{0.031}&$~~~$
            \\
            &3D-FCNN{\cite{mei2017hyperspectral}}$~$&$~~~${0.623}&$~~~$
            \\	
            &GDRRN{\cite{li2018single}}&$~~~${0.551}&$~~~$
            \\
            &DHP{\cite{sidorov2019deep}}&$~~~${704.879}&$~~~$
            \\
            &3D-GAN{\cite{li2020hyperspectral}}&$~~~${0.638}&$~~~$
            \\
            &SSPSR{\cite{jiang2020learning}}$~$&$~~~$0.659&$~~~$
            \\
            &ADMM-ADAM SR{\cite{lin2022single}}$~$&$~~~$156.305&$~~~$
            \\
            &Proposed{}$~$&$~~~${0.660}&$~~~$
            \\
            \hline
		\end{tabular}}
	\end{table}

	\section{Conclusion}\label{sec:conclusion}
    In this paper, we have presented Hyper-Restormer for remote sensing HSI restoration. We use Spectral  Attention Block, Spatial Attention Block, and LLFF to compose the LSS Transformer Block. The former attention blocks extract long-range dependencies from spectral and spatial domains, while the latter enhances local context information through a lightweight computation. Then, the LSS Transformer Block forms the SLSST through a novel low-rank model architecture, reducing the extensive computational cost required by self-attention. Finally, multiple SLSSTs are cascaded to form Hyper-Restormer, progressively enhancing the quality of remote sensing HSI restoration. Extensive experiments were conducted on HSI restoration tasks, including denoising, inpainting, and super-resolution, demonstrating that the Hyper-Restormer achieves state-of-the-art performance.

	\bibliographystyle{IEEEtran}
	\bibliography{ref}

\begin{IEEEbiography}[{\resizebox{0.9in}{!}{\includegraphics[width=1in,height=1.25in,clip,keepaspectratio]{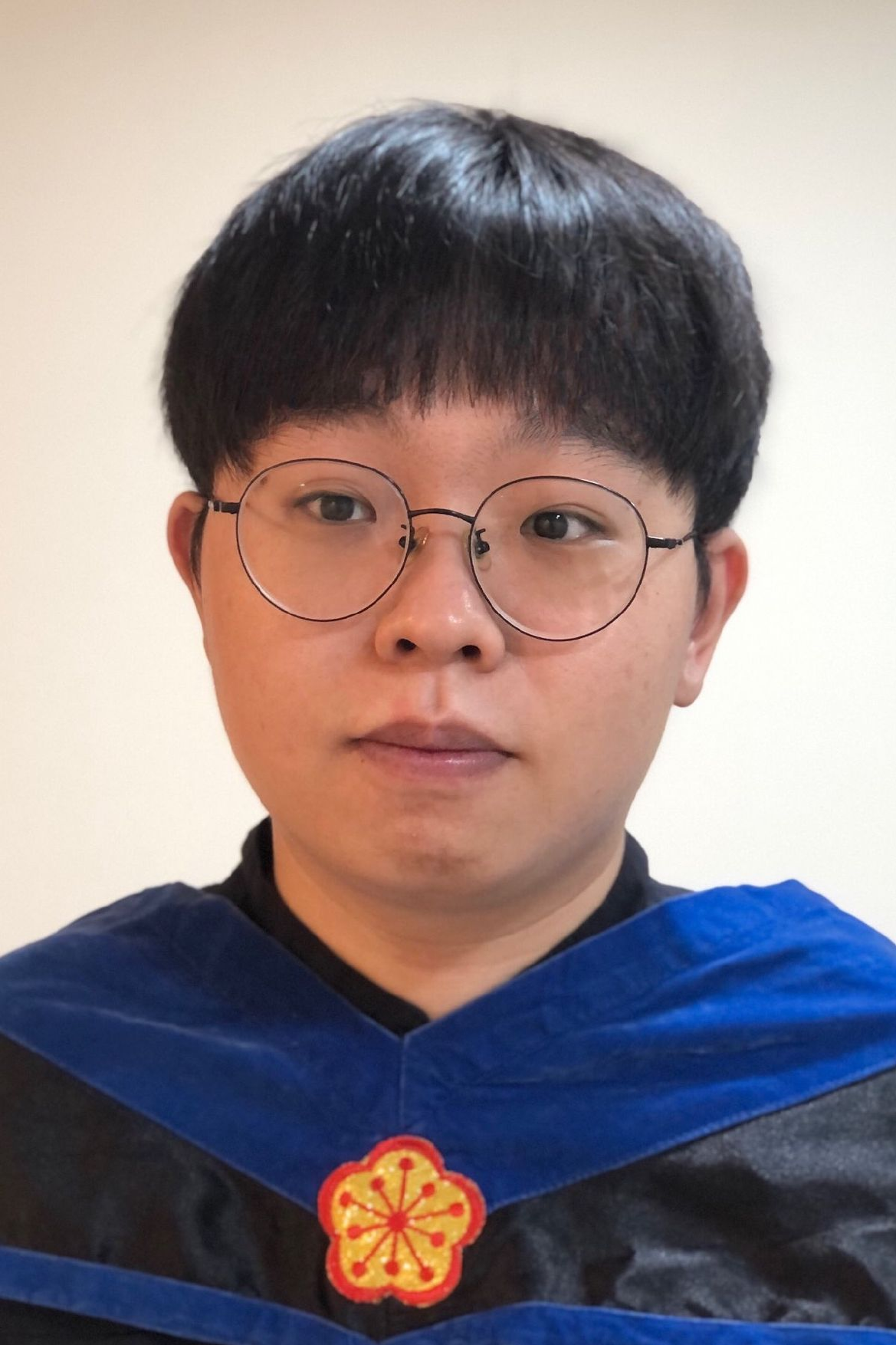}}}]
		{\bf Yo-Yu Lai}
		received his B.S. degree from the Department of Engineering Science, National Cheng Kung University, Taiwan, Taiwan, in 2021.
		
		He is currently a graduate student with Intelligent Hyperspectral Computing Laboratory, Institute of Computer and Communication Engineering, National Cheng Kung University, Taiwan. 
		His research interests include deep learning, convex optimization, hyperspectral imaging, and adversarial defense.
		He was a recipient of the Outstanding Paper Award from the Chinese Image Processing and  Pattern Recognition Society (IPPR) Conference on Computer Vision, Graphics, and Image Processing (CVGIP), in 2021.
	\end{IEEEbiography}

 \begin{IEEEbiography}[{\resizebox{0.9in}{!}{\includegraphics[width=1in,height=1.25in,clip,keepaspectratio]{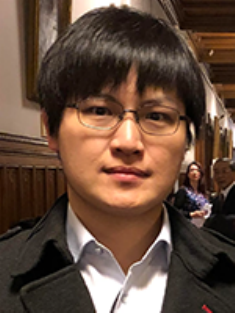}}}]
	{\bf Chia-Hsiang Lin}
    (S'10-M'18)
    received the B.S. degree in electrical engineering and the Ph.D. degree in communications engineering from National Tsing Hua University (NTHU), Taiwan, in 2010 and 2016, respectively.
    From 2015 to 2016, he was a Visiting Student of Virginia Tech,
    Arlington, VA, USA.
    
    He is currently an Associate Professor with the Department of Electrical Engineering, and also with 
    the Miin Wu School of Computing,
    National Cheng Kung University (NCKU), Taiwan.
    Before joining NCKU, he held research positions with The Chinese University of Hong Kong, HK (2014 and 2017), 
    NTHU (2016-2017), 
    and the University of Lisbon (ULisboa), Lisbon, Portugal (2017-2018).
    He was an Assistant Professor with the Center for Space and Remote Sensing Research, National Central University, Taiwan, in 2018, and a Visiting Professor with ULisboa, in 2019.
    His research interests include network science, 
    quantum computing,
    convex geometry and optimization, blind signal processing, and imaging science.
    
    Dr. Lin received the Emerging Young Scholar Award from National Science and Technology Council (NSTC), in 2023,
    the Future Technology Award from NSTC, in 2022,
    the Outstanding Youth Electrical Engineer Award from The Chinese Institute of Electrical Engineering (CIEE), in 2022,
    the Best Young Professional Member Award from IEEE Tainan Section, in 2021,
    the Prize Paper Award from IEEE Geoscience and Remote Sensing Society (GRS-S), in 2020, 
    the Top Performance Award from Social Media Prediction Challenge at ACM Multimedia, in 2020,
    and The 3rd Place from AIM Real World Super-Resolution Challenge at IEEE International Conference on Computer Vision (ICCV), in 2019. 
    He received the Ministry of Science and Technology (MOST) Young Scholar Fellowship, together with the EINSTEIN Grant Award, from 2018 to 2023.
    In 2016, he was a recipient of the Outstanding Doctoral Dissertation Award from the Chinese Image Processing and Pattern Recognition Society and the Best Doctoral Dissertation Award from the IEEE GRS-S.

\end{IEEEbiography}

	\begin{IEEEbiography}[{\resizebox{1in}{!}{\includegraphics[width=1in,height=1.25in,clip,keepaspectratio]{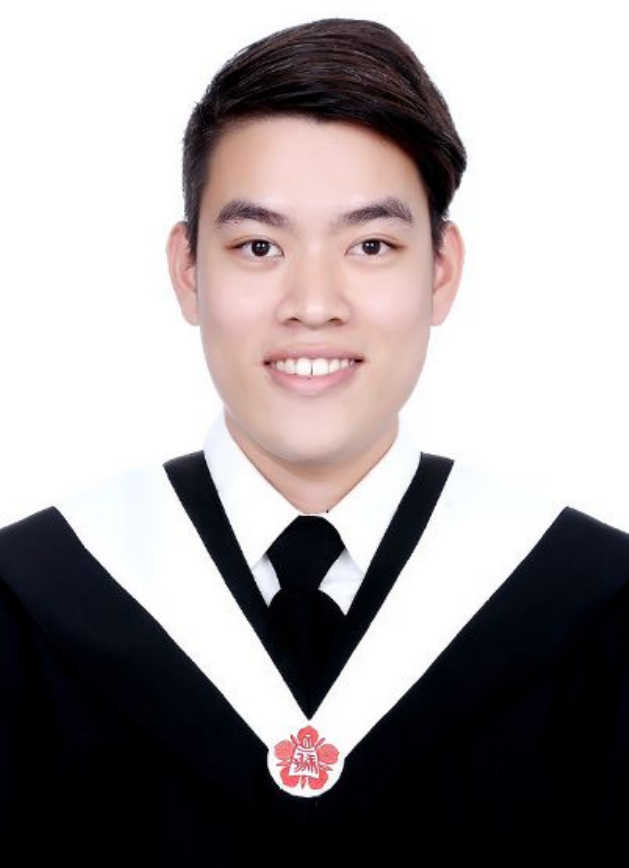}}}]
		{\bf Zi-Chao Leng}
		received his B.S. degree from the Department of Electronic Engineering, National Cheng Kung University, Taiwan, in 2021.
		
		He is currently a Ph.D. student with Intelligent Hyperspectral Computing Laboratory, Institute of Computer and Communication Engineering, National Cheng Kung University, Taiwan. 
		His research interests include deep learning, convex optimization, hyperspectral imaging, and medical imaging.
	\end{IEEEbiography}

\end{document}